\PassOptionsToPackage{table}{xcolor}
\documentclass[letterpaper,twocolumn,10pt]{article}
\usepackage{usenix2019_v3}
\usepackage{caption}
\usepackage{framed}
\usepackage{multirow}
\usepackage{booktabs}
\usepackage{ifthen}
\usepackage{tablefootnote}
\usepackage{xspace}
\usepackage[T1]{fontenc}
\usepackage{enumerate}
\usepackage[linesnumbered, ruled, vlined]{algorithm2e}
\usepackage{array,graphicx}
\usepackage{float}
\usepackage{balance}
\usepackage{tikz}
\usepackage{calc}
\usepackage{subfigure}
\usepackage{amsfonts}
\usepackage[table]{xcolor}
\usepackage{xspace}
\usepackage{hyperref,endnotes}
\usepackage{array}
\usepackage{makecell}
\usepackage[misc]{ifsym}
\usepackage{svg}
\usepackage{graphicx}
\usepackage{tabularx}
\usepackage{adjustbox}
\usepackage{ragged2e}
\usepackage{cite}
\usepackage{pifont}
\usepackage{amssymb}
\usepackage{url}
\usepackage[numbers,sort]{natbib}
\usepackage[most]{tcolorbox}
\usepackage{longtable}
\usepackage{helvet}
\usepackage{caption}
\usepackage{subcaption}
\usepackage{subfigure}
\usepackage{ulem}
\usepackage{pifont}

\lstdefinelanguage{Solidity}{
    morekeywords={
        public, Exchange, function, memory, private, delegatecall, internal, new, while, if, else
    },
    morekeywords=[2]{address, bytes, uint256, super},
    sensitive=true,
    morecomment=[l]{//},
    morecomment=[s]{/*}{*/},
    morestring=[b]",
}

\lstdefinestyle{mystyle}{
    language=Java,
    basicstyle=\ttfamily\small,
    keywordstyle=\color{blue},
    stringstyle=\color{red},
    commentstyle=\color{green},
    numberstyle=\tiny\color{gray},
    stepnumber=1,
    numbersep=5pt,
    tabsize=2,
    breaklines=true,
    showtabs=false,
    showspaces=false,
    showstringspaces=false,
    identifierstyle=\ttfamily,
    frame=single,
    rulecolor=\color{black},
    title=\lstname,
    breakindent=0em,
    breakatwhitespace=true,
    prebreak=\raisebox{0ex}[0ex][0ex]{\ensuremath{\color{blue}\hookrightarrow}},
    postbreak=\raisebox{0ex}[0ex][0ex]{\ensuremath{\color{blue}\hookrightarrow}}
}

\definecolor{keywordcolor}{rgb}{0.13, 0.13, 1.0}
\definecolor{typecolor}{rgb}{0.33, 0.10, 0.62} 
\definecolor{commentcolor}{rgb}{0.5, 0.5, 0.5}
\definecolor{stringcolor}{rgb}{0.58, 0.0, 0.82}
\definecolor{bgcolor}{rgb}{0.95, 0.95, 0.95}

\AtEndPreamble{
	\usepackage{hyperref}
	\hypersetup{
		colorlinks = true,
		linkcolor = purple,
		anchorcolor = purple,
		citecolor = purple,
		filecolor = purple,
		urlcolor = black
	}
}

\newcommand{\tool}{SMARTCAT}
\newcommand{\sx}{\textit{Extractor}}
\newcommand{\sy}{\textit{Builder}}
\newcommand{\sz}{\textit{Detector}}

\begin{document}

\title{Following Devils' Footprint: Towards Real-time Detection of \\Price Manipulation Attacks}

\author{
{\rm Bosi Zhang$^{1}$, Ningyu He$^{2}$\ddag, Xiaohui Hu$^{1}$, Kai Ma$^{1}$ , Haoyu Wang$^{1}$\ddag}
\\
$^{1}$ Huazhong University of Science and Technology
$^{2}$ The Hong Kong Polytechnic University\\
\rm \ddag Co-corresponding authors: ningyu.he@polyu.edu.hk, haoyuwang@hust.edu.cn
}

\maketitle

\begin{abstract}
Price manipulation attack is one of the notorious threats in decentralized finance (DeFi) applications, which allows attackers to exchange tokens at an extensively deviated price from the market. Existing efforts usually rely on \textit{reactive} methods to identify such kind of attacks after they have happened, \textit{e.g.,} detecting attack transactions in the \textit{post-attack stage}, which cannot mitigate or prevent price manipulation attacks timely.
From the perspective of attackers, they usually need to deploy attack contracts in the \textit{pre-attack stage}. Thus, if we can identify these attack contracts in a \textit{proactive} manner, we can raise alarms and mitigate the threats. 
With the core idea in mind, in this work, we shift our attention from the victims to the attackers.
Specifically, we propose {\tool}, a novel approach for identifying price manipulation attacks in the \textit{pre-attack stage} proactively. 
For generality, it conducts analysis on bytecode and does not require any source code and transaction data.
For accuracy, it depicts the control- and data-flow dependency relationships among function calls into a token flow graph.
For scalability, it filters out those suspicious paths, in which it conducts inter-contract analysis as necessary.
To this end, {\tool} can pinpoint attacks in real time once they have been deployed on a chain.
The evaluation results illustrate that {\tool} significantly outperforms existing baselines with 91.6\% recall and $\sim$100\% precision. Moreover, {\tool} also uncovers 616 attack contracts in-the-wild, accounting for \$9.25M financial losses, with only 19 cases publicly reported. By applying {\tool} as a real-time detector in Ethereum and Binance Smart Chain, it has raised 14 alarms 99 seconds after the corresponding deployment on average. These attacks have already led to \$641K financial losses, and seven of them are still waiting for their ripe time.
\end{abstract}
\section{Introduction}

Since the emergence of Ethereum, smart contracts have been its killer application, \textit{i.e.,} the feature distinguishes it from old-school Bitcoin~\cite{nakamoto2008bitcoin}.
By decoupling complex interactions into different smart contracts, developers are able to build Decentralized Applications (DApps)~\cite{raval2016decentralized}, and even Decentralized Finance (DeFi)~\cite{zetzsche2020decentralized}, which specifically provides financial services, like lending, exchanging, and even insurance, in a decentralized manner~\cite{shah2023systematic}.
In 2024, the total value locked, one of the critical metrics to reflect the prosperity, in DeFi protocols has surged to over \$90 billion~\cite{biance_re}.

Unfortunately, due to the anonymity and immutability of Ethereum smart contracts, numerous DeFi projects are exploited by unidentifiable accounts, leading to \$473 million financial losses in 2024~\cite{coindesk}.
Among all vulnerabilities in DeFi protocols, \textit{price manipulation} must be one of the most notorious ones~\cite{zhang2023demystifying}. In short, attackers can obtain massive profits from token transfers or exchanges at a price far from the market's normal fluctuation.
Various reasons could finally lead to a price manipulation attack, \textit{e.g.,} incorrect slippage settings, unprotected public functions, and reliance on untrusted price oracles~\cite{mo2023toward}. Furthermore, cunning attackers would take advantage of the Flashloan mechanism~\cite{qin2021attacking} to conduct exploitation to ensure that the whole attack transaction can be rolled back promptly if any condition is not met.

Existing studies against price manipulation are in two forms, \textit{i.e.,} either \textit{reactively identifying attacks in the post-attack stage according to transaction data} or \textit{detecting if there are price manipulation vulnerabilities in DeFi protocols}.
As for the former one, DeFiRanger~\cite{wu2023defiranger} constructs a cash flow tree from transaction traces, while DeFiGuard~\cite{wang2024defiguard} extracts behavioral patterns from transactions and combines them with a graph neural network. However, neither tool can mitigate such attacks proactively.
As for the latter form, FlashSyn~\cite{chen2024flashsyn} uses a numerical approximation to synthesize transactions for exploiting Flashloan-based price manipulation vulnerabilities, and DeFiTainter~\cite{kong2023defitainter} detects vulnerabilities in DeFi projects using generic rules.
However, most price manipulation attacks stem from poorly designed contract business logic, which have to be manually modeled one by one in these methods, significantly impacting their scalability.

In other words, \textit{current work cannot mitigate or prevent price manipulation attacks effectively and timely}.
Therefore, standing from the attackers' perspective, in this work, we \textit{proactively detect these attacks in the pre-attack stage}. We focus on newly deployed Ethereum contracts and uncover if they possess such malicious intent.
However, two main challenges arise.
On the one hand, in Ethereum, only 2\% contracts are open-sourced~\cite{percent}. Furthermore, to cover their malicious intent, such attack contracts typically avoid releasing their implementations. We have to precisely recover their behavioral semantics without introducing too many false positives and negatives.
On the other hand, currently, there are around 39K newly deployed contracts a day in Ethereum~\cite{daily}. We have to precisely identify these attack contracts out of numerous benign ones. Moreover, rational attackers must initiate attacks once everything is settled down, which also requires the timeliness of our detecting methods.

\textbf{This Work.}
In this work, we propose {\tool}, a static analysis framework to identify price manipulation attack contracts.
To recover the behavioral semantics, it extracts the callee address and invoked function for each function call. We also propose a \textit{data-flow-based heuristic arguments recovery algorithm} to recover the arguments for these function calls.
Based on the inter-procedural control flow graph (ICFG), we construct \textit{cross-function callsite graph} (xFCG) and \textit{token flow graph} (TFG) to depict the control- and data-flow dependency relations among function calls.
To enable efficient analysis, we propose a \textit{sensitive path filtering method} to selectively conduct cross-contract analysis on the TFG. Additionally, we formalize a set of rules to characterize two types of price manipulation attack behaviors in a sound and precise manner.

In the evaluation, we construct a ground truth dataset consisting of 84 price manipulation attack contracts and 8,000 benign contracts. The results show that {\tool} can correctly identify 77 out of the 84 attack contracts, with a negligible false positive rate, extensively outperforming the existing three available baselines.
We then apply {\tool} on a large-scale dataset with over 77K real-world contracts. It identifies 616 price manipulation attacks, where only 19 cases were reported previously, causing \$9.25M financial losses in total.
Furthermore, to evaluate its timeliness, we deploy {\tool} on Ethereum and Binance Smart Chain for 50 days. In total, {\tool} has raised 14 alarms 99 seconds after the corresponding contract deployment on average. These attacks have led to \$641K financial losses already, and seven contracts are still waiting for the ripe time.

We summarize our contributions as follows:

\begin{itemize}
    \item We propose {\tool} for identifying price manipulation attack contracts through anomalous token flow. To the best of our knowledge, this is the first work to identify such attack contracts on the bytecode level.
    \item Based on an extensively constructed ground truth dataset, the precision and recall of {\tool} are $\sim$100\% and 91.6\%, respectively, significantly outperforming existing available baselines and demonstrating robustness against obfuscation techniques.
    \item {\tool} have identified 616 price manipulation attack contracts out of  770K real-world deployed contracts, where only 19 were reported publicly. {\tool} can analyze 99\% cases within 40.6 seconds.
    \item As a real-time detector, {\tool} has raised alarms 14 times in Ethereum and BSC. These attacks have led to \$641K financial losses already.
\end{itemize}

\section{Background}

\subsection{Smart Contracts \& Decentralized Finance}
\label{sec:smart}
Ethereum smart contracts are self-executing programs typically implemented in high-level languages like Solidity~\cite{dannen2017introducing}. They are compiled into bytecode, deployed, and executed in the Ethereum Virtual Machine (EVM)~\cite{hirai2017defining}, a stack-based virtual machine.
The immutable nature of the blockchain makes smart contracts susceptible to vulnerabilities once deployed. To address this, smart contracts often utilize proxy mechanisms~\cite{eip1967} for upgrading their logic or fixing bugs.
Smart contracts use three data structures to maintain data: \textit{Memory}, \textit{Storage}, and \textit{Calldata}. \textit{Memory} holds temporary data required during function execution, \textit{Storage} stores data permanently on-chain, and \textit{Calldata} contains read-only function arguments passed from external calls initiated by opcodes like CALL and DELEGATECALL.

DeFi, an emerging financial ecosystem built on smart contracts, showcases immense potential.
We highlight some widely-adopted DeFi protocols in the following.

\noindent
\textbf{Token}. Except for the native token in Ethereum, accounts are allowed to create and issue tokens by implementing some standards, \textit{e.g.,} ERC-20~\cite{erc20} and ERC-721~\cite{erc721}.
Moreover, DeFi widely leverages \textit{stablecoins} and \textit{liquidity provider (LP) tokens}. Specifically, stablecoins can ensure price stability by anchoring specific currencies, like USD, offering a reliable medium for value exchange. LP tokens are issued by DeFi projects to represent someone's shares and benefits.

\noindent
\textbf{Decentralized Exchange}. Decentralized exchange (DEX) uses smart contracts for token exchange, ensuring transparency, openness, and trustlessness. Lots of well-known DEXes exist in Ethereum, like Uniswap~\cite{uniswap}.
DEX typically employs the \textit{automated market maker} (AMM) mechanism, dynamically adjusting token prices based on the token reserve in liquidity pools.
Users can deposit tokens into DEXes to provide liquidity and earn interest in an LP token form.

\noindent
\textbf{Flashloan}. Flashloan is a form of uncollateralized loan that must be taken and 
Protocols such as Aave~\cite{aave} and Uniswap~\cite{uniswap} provide flashloan services, allowing users to borrow large amounts without upfront collateral. 
Though flashloan provides convenience, it has also led to various attack incidents in recent years~\cite{wang2020towards, qin2021attacking}.

\subsection{Price Manipulation}
\label{sec:pm}
Price manipulation in DeFi is basically achieved by exploiting how DEXes calculate token prices.
Typically, for a token in a liquidity pool, the lower the supply, the higher the price. When the supply balance between tokens is disrupted, price deviations may occur. 
The existence of flashloan further worsens this issue by enabling someone to borrow large amounts of assets without collateral, \textit{i.e.,} causing rapid and significant price fluctuation before the loan is repaid.
Based on how the victim contract is affected, it is divided into \textit{direct price manipulation (DPM)} and \textit{indirect price manipulation (IPM)}.

\begin{figure}[t]
    \centering
    \subfigure[DPM in \textit{ElephantStatus}.]{
        \centering
        \includegraphics[width=0.45\columnwidth]{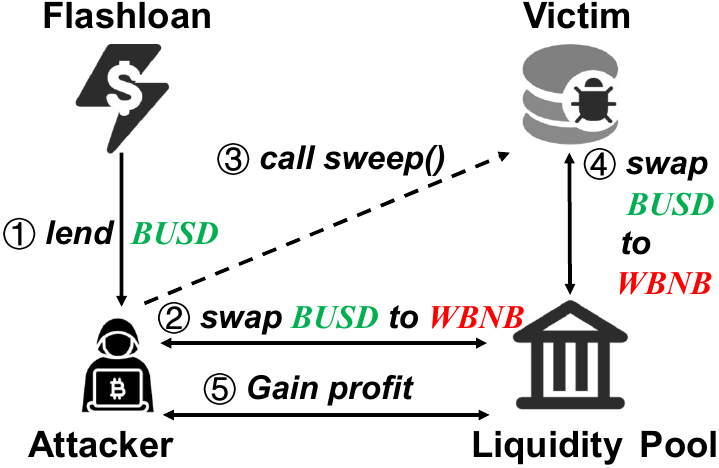}
        \label{fig:elephant}
    }
    \hspace{0.02\columnwidth} 
    \subfigure[IPM in \textit{Cheese Bank}.]{
        \centering
        \includegraphics[width=0.45\columnwidth]{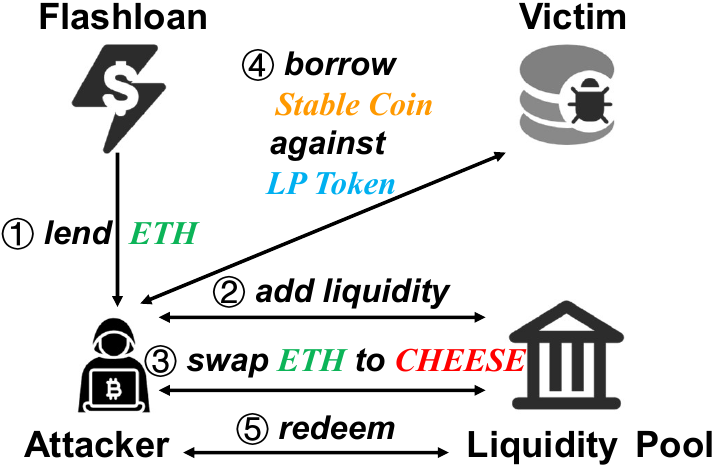}
        \label{fig:cheese}
    }
    \vspace{-0.2in}
    \caption{Two types of price manipulation.}
    \vspace{-0.1in}
    \label{fig:pm}
\end{figure}

\noindent
\textbf{Direct Price Manipulation (DPM).}
\label{sec:dpm}
By exploiting calculation errors or access control issues within the victim contract, attackers can directly perform price manipulation in the liquidity pools of DEXes. 
This typically involves three steps: \textbf{i)} the attacker uses a large amount of token \texttt{A} to exchange for token \texttt{B} in the liquidity pool, causing the price of token \texttt{B} to increase; \textbf{ii)} the attacker further decreases the liquidity of token \texttt{B} by exploiting the victim contract, \textit{e.g.,} let the victim purchase a large amount of token \texttt{B}, driving up the price of token \texttt{B}; and \textbf{iii)} the attacker sells the acquired token \texttt{B} and profits from the price manipulation.

Figure~\ref{fig:elephant} illustrates a real-world example of direct price manipulation, where \textit{ElephantStatus} was exploited on Dec. 6th, 2023, resulting in approximately \$165K financial losses~\cite{elephant}.
As we can see, the attacker borrows a large amount of BUSD tokens via flashloan (step \ding{172}), which are then swapped to WBNB tokens (step \ding{173}). The attacker then calls \texttt{sweep} in the victim contract (step \ding{174}), which transfers a large amount of BUSD into the liquidity pool and withdraws an equivalent value of WBNB from the pool (step \ding{175}). This operation artificially inflates the price of WBNB in the liquidity pool.
Finally, the attacker swaps the previously acquired WBNB back to BUSD (step \ding{176}), a stablecoin, obtaining profits by such a direct price manipulation.
Note that,  there is a variant of DPM, where attackers can also directly leverage flashloans to significantly impact the asset reserves in the liquidity pool, subsequently profiting from the price fluctuations. In this situation, the victim is the liquidity pool itself.

\noindent
\textbf{Indirect Price Manipulation (IPM).}
\label{sec:ipm}
Instead of directly exploiting the victim contract, attackers conduct indirect price manipulation by disturbing token prices in DEXes, which are adopted as price oracles by victim contracts.
Specifically, it generally consists of three steps: \textbf{i)} the attacker intentionally creates an imbalance in the token reserves of a liquidity pool; \textbf{ii)} the attacker interacts with the victim contract, which calculates the token prices in real-time based on the oracle exposed by the liquidity pool. To this end, the attacker can sell or stake tokens to the victim contract at an inflated price; and \textbf{iii)} the attacker restores the balance of the liquidity pool.

Figure~\ref{fig:cheese} illustrates a concrete example of an indirect price manipulation, which occurred on Nov. 6th, 2020, leading to financial losses of Cheese Bank estimated at \$3.3M~\cite{cheese}.
Specifically, the attacker first borrows a large amount of ETH via flashloan (step \ding{172}). Then, the attacker deposits ETH and CHEESE tokens to the liquidity pool in exchange for the corresponding number of LP tokens (step \ding{173}).
The attacker further swaps a large amount of ETH for CHEESE tokens (step \ding{174}).
Since \textit{Cheese Bank} calculates the price of LP tokens based on the amount of ETH in the liquidity pool, through a legitimate external call, the attacker drains the victim contract, which is tricked into thinking that the price of LP tokens is extremely high (step \ding{175}).
Finally, the attacker redeems ETH from the liquidity pool (step \ding{176}).

\subsection{Threat Model}

As for conducting price manipulation attacks, compared to ordinary accounts in Ethereum, attackers have no other extra privileges. All attack logic is embedded in the deployed contract, and the attack is launched by initiating a transaction.
As illustrated in \S\ref{sec:pm}, both liquidity pools and contracts that have the ability to interact with the pools can be potential victims.
\textit{Leveraging the time window between the attack contract deployment and the attack launch is critical.} In the real world, attackers may delay the attack until certain conditions are met or until the profit is maximized. 
\section{Motivation}
\label{sec:acb}

In this section, we first demonstrate what a price manipulation attack contract looks like. Then, we summarize the challenges in identifying these contracts on the contract bytecode level and illustrate our solution in a high-level manner.

\subsection{Motivating Example}
\label{sec:moti_eg}
\textit{ULME} project has been attacked in a DPM way on Oct. 25th, 2022, suffering from \$50K financial losses~\cite{ulme}. Listing~\ref{lst:motivating} illustrates an attacker-defined private function, which is invoked by the callback function once the attacker takes out a flashloan. The function is shown in a simplified and decompiled way.
As we can see, the attacker first extracts the addresses of BUSDT and ULME token from \textit{Storage}, stores them into a newly initialized array, and then calls the token swap function in the liquidity pool (L3 -- L7).\footnote{L3 refers to the 3rd line, we adopt this notation in the following.}
In this example, the attacker exchanges BUSDT for ULME token. Subsequently, the attacker iterates \texttt{array\_9} (L8), consisting of pre-identified users who have approved the BUSDT contract, to filter out those with non-zero balances by calling \texttt{allowance()} and \texttt{balance()} (L9 -- L13).
Unlike IPM, where the victim indirectly relies on the liquidity pool as the oracle to calculate the price of ULME, the attacker invokes \texttt{buyMiner()} of the ULME contract. This step \textit{directly} uses the victim's BUSDT to swap out ULME from the liquidity pool, sharply reducing its ULME supply and further destabilizing the pool's state.
The attacker then exchanges ULME back for BUSDT token (L17 -- L21). Such a swap obtains a large amount of BUSDT at an unfair price, profiting from the imbalance in the pool.

\begin{lstlisting}[caption={Attack contract against \textit{ULME}.}, label=lst:motivating]
function 0x9e1() private {
    if (stor_a) {
        v2 = new address[](2);
        v2[0] = stor_2; v2[1] = stor_5;
        v4 = new address[](v2.length);
    // Step_1: exchange BUSDT for ULME
        v11 = stor_4.swapExactTokensForXXX(stor_a, 0, v4, address(this), block.timestamp + 1000, v12, stor_2).gas(msg.gas);}
    while (v13 < array_9.length){
        v15, v16 = stor_2.allowance(address(array_9[v13]), stor_5).gas(msg.gas);
        if (v16) {
            v17, v18 = stor_2.balanceOf(address(array_9[v13])).gas(msg.gas);
            v19 = v20 = v18 > 0;
            if (v19) {
                require(bool(stor_5.code.size));
    // Step_2: manipulate users into swapping out their ULME
                v22 = stor_5.buyMiner(address(array_9[v13]), v18 * 100/110 + ~0).gas(msg.gas);}}}
    v26 = new address[](2);
    v26[0] = stor_5; v26[1] = stor_2;
    v28 = new address[](v26.length);
    // Step_3: exchange ULME for BUSDT at an unfair price
    v35 = stor_4.swapExactTokensForXXX(v25, 0, v28, address(this), block.timestamp + 1000, v12, stor_5).gas(msg.gas);
}
\end{lstlisting}

\subsection{Challenges \& Solution}
Through this example, we can find that attackers will leave traces in their contracts, including token manipulation, the use of Flashloan services, and interactions with liquidity pools. Due to the transparency of the blockchain, these features can be obtained at the time of contract deployment, allowing us to promptly raise alarms for suspicious attack contracts.
However, two key challenges need to be addressed.

\noindent
\textbf{Challenge 1: Unclear semantics.}
Attack contracts are typically close-sourced to conceal their malicious intent, limiting analysis to the bytecode level. To recover semantics, existing bytecode-based tools either elevate the bytecode to an intermediate representation (IR)~\cite{grech2019gigahorse} or adopt static analysis methods, like symbolic execution~\cite{mythril}.
However, on the one hand, the obtained IR is limited to the contract itself and does not provide cross-contract semantics. On the other hand, the complexity of cross-contract calls and state dependencies between contracts can lead to path explosion during symbolic execution.
As shown in Listing~\ref{lst:motivating}, recovering and identifying semantics at both intra- and inter-contract function calls is crucial for accurately determining the contract's behavior.

\noindent
\textbf{Challenge 2: Scalability issue.}
Detecting price manipulation attack contracts requires exploring the paths corresponding to conducting attacks. In Listing~\ref{lst:motivating}, we only illustrate the function that performs attacks and omits other auxiliary functions, which could introduce complexity through loops, conditional branches, and even inter-contract calls.
Thus, we have to thoroughly analyze all defined functions within the contract and effectively identify the attack path among numerous paths.
Furthermore, as timeliness is crucial for avoiding under-reporting, we must efficiently explore paths and minimize interference from irrelevant ones.

\noindent
\textbf{Our Solution:}
Against \textbf{Challenge 1}, we first extract both callee addresses and invoked functions from intra- and inter-contract function calls, where we propose a fine-grained argument recovery algorithm to retrieve concrete values of their arguments.
Furthermore, we take advantage of the function signature database and heuristic rules to capture the operational semantics of all function calls.
As for \textbf{Challenge 2}, rather than relying on machine learning or heuristic rule based methods, we adopt a formal approach to model price manipulation attack behaviors. To improve efficiency, we filter out all suspicious sensitive paths based on characteristics of DeFi attacks and limit the scope of cross-contract analysis.

\section{Methodology}
\label{sec:methodology}

\begin{figure}[t]
    \captionsetup{justification=centering}
    \centering
    \includegraphics[width=\columnwidth]{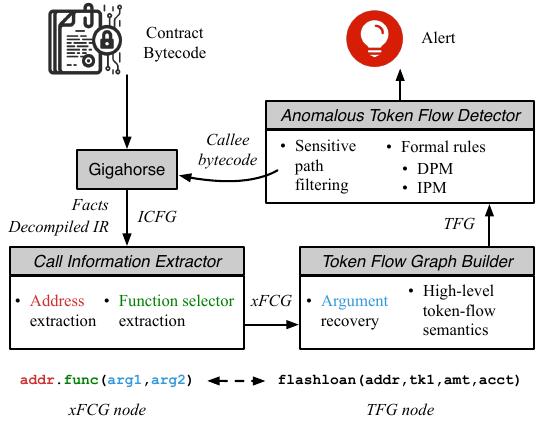}
    \vspace{-0.2in}
    \caption{Workflow and architecture of {\tool}.}
    \vspace{-0.1in}
    \label{fig:workflow}
\end{figure}

To identify attack contracts, we propose {\tool}, whose workflow and architecture are shown in Figure~\ref{fig:workflow}. As we can see, {\tool} is composed of three modules, \textit{i.e.,} \textit{Call Information Extractor} (short as {\sx}), \textit{Token Flow Graph Builder} (short as {\sy}) and \textit{Anomalous Token Flow Detector} (short as {\sz}).
Specifically, {\sx} is built upon Gigahorse~\cite{grech2019gigahorse, grech2022elipmoc}, a well-known tool that can decompile and produce intermediate representation (IR) in three-address code format. {\sx} takes the contract bytecode as input and decompiles it to obtain the IR and control flow information on the function level.
To address \textbf{Challenge 1}, based on the inter-procedural control flow graph, {\sx} firstly constructs a cross-function callsite graph (xFCG) consisting of nodes, each of which encompasses information of the callsite, callee address, and invoked function signature. The generated xFCG will be transmitted to {\sy}, which employs a data-flow-based heuristic arguments recovery algorithm to retrieve arguments' values of invoked functions. By combining the token-flow-related semantics of involved functions, {\sy} further builds a token flow graph (TFG).
To address \textbf{Challenge 2}, {\sz} filters out and traverses those suspicious sensitive paths in the TFG. By adopting a set of formal detecting rules, {\sz} can finally identify price manipulation attack contracts in an effective and efficient manner.

\subsection{Call Information Extractor}
\label{sec:extract}
As illustrated in \S\ref{sec:moti_eg}, invoking functions in both intra- and inter-contract manner is necessary for profiting from conducting price manipulation, making the extraction of call information important.
To achieve this, we leverage Gigahorse~\cite{grech2019gigahorse} and introduce a \textit{Cross-Function Callsite Graph (xFCG)} based on the inter-procedural control flow graph to efficiently depict and extract this information.

\subsubsection{Address \& Function Extraction}
\label{method:extractor:address}
\label{method:extractor:function}
Extracting callee addresses for function calls is the prerequisite for the following analysis.
In Ethereum, callee addresses can be specified in three different data structures: \textit{Calldata}, \textit{Memory}, and \textit{Storage} (see \S{\ref{sec:smart}}).
Since \textit{Calldata} is provided at runtime, which is unpredictable, we assign a placeholder in the form of \texttt{calldata\_0xN}, where \texttt{0xN} corresponds to its offset in \textit{Calldata}. For the other two cases, we leverage the \textit{facts} generated by Gigahorse based on the decompiled code. Specifically, if the callee address is stored in \textit{Memory}, we either directly extract the address (when hard-encoded in bytecode) or assign a placeholder, the same as \textit{Calldata}. Otherwise, \textit{i.e.,} the callee address is stored in \textit{Storage}, we extract the hard-encoded callee address from the facts or retrieve it from a slot number via the \texttt{getStorageAt()} API.
Note that some contracts adopt the proxy design pattern, where the actual function resides in another contract that the proxy points to. Since EIP-1967 is the most widely-adopted proxy standard in Ethereum~\cite{eip1967}, whose \textit{Storage} structure is fixed, we heuristically investigate if the EIP-1967-specific slot exists~\cite{openeip}. If it is, we take the current contract as a proxy and extract the real callee address from the slot.

\begin{figure}[t] 
\centering
\includegraphics[width=\columnwidth]{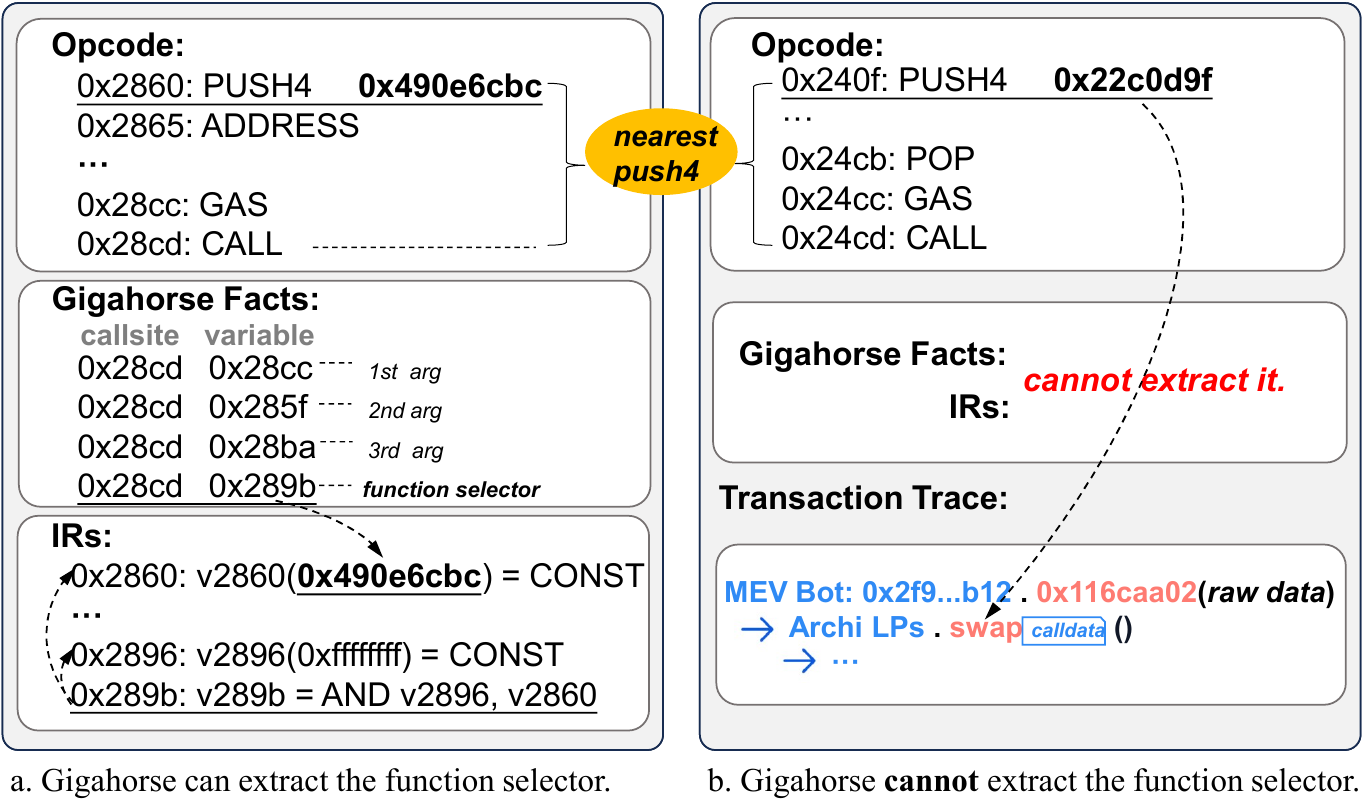} 
\vspace{-0.2in}
\caption{Extract function signature by Gigahorse and our heuristic method.} 
\vspace{-0.1in}
\label{fig:selector} 
\end{figure}

Instead of the callee address, we also need to identify the invoked function to recover the developers' intent. In Ethereum contract bytecode, each function call consists of the function signature and its arguments.
As shown in Figure~\ref{fig:selector}{\color{purple}(a)}, we leverage the facts generated by Gigahorse to directly extract the 4-byte function signature. However, Gigahorse may fail to generate such facts when branch-dependent \textit{Memory} read and write operations obscure the exact \textit{Memory} layout~\cite{grech2019gigahorse, zhang2023bian}.To address this, we propose a heuristic method: backtracing the control flow to extract the operand of the most recent \texttt{PUSH4} opcode before the function call, treating it as the function signature.
As shown in Figure~\ref{fig:selector}{\color{purple}(b)}, in this case, Gigahorse fails in this MEV bot contract due to frequent branch jumps~\cite{akmev}, whereas our method correctly identifies the callee function as \texttt{swap()}, confirmed by the transaction trace.

\subsubsection{Cross-Function Callsite Graph Construction}
\label{method:xcsg}
To illustrate relationships among function calls, based on the inter-procedural control flow graph (ICFG), we propose and build the cross-function callsite graph (xFCG).
Specifically, we only consider the basic blocks that contain function calls. We adopt a triplet to refer to each of them, defined as:
$$<callsite,callee\_address,function\_selector>$$
where $callsite$ is the offset of the call-related opcode, and $callee\_address$ and $function\_selector$ refer to the corresponding information extracted in \S\ref{method:extractor:address}.
For those basic blocks without function calls, we directly remove them and link their preceding and successive blocks.
Note that both intra-contract and inter-contract function calls are taken into consideration in building xFCG, \textit{i.e.,} the $callee\_address$ can be the address of the current or another contract.
If an inter-contract function call happens, a recursive inter-contract xFCG build will only happen on those nodes located in suspicious paths, whose selection rules will be detailed in \S\ref{sec:spf}.

\subsection{Token Flow Graph Builder}
\label{sec:tfg_g}
Identifying the value of arguments in function calls can recover the original intents more precisely. Thus, we propose a \textit{data-flow-based heuristic arguments recovery algorithm}.
Moreover, we focus on five types of token-flow related semantics, \textit{e.g.,} swap token and add liquidity, which are combined on nodes in xFCG to construct the \textit{token flow graph} (TFG).

\subsubsection{Argument Recovery}
\label{lab:argu}
The xFCG we built can only reflect the control flow dependency relationships among function calls. To more effectively address \textbf{Challenge 1}, we will also recover data flow dependency relationships, whose very first step is to recover the values or variables associated with arguments in function calls.
In EVM bytecode, call instructions do not explicitly declare them. Instead, it points a piece of bytes in \textit{Memory} through an offset and length.
While we can obtain parameter information through Gigahorse's facts, as introduced in \S\ref{method:extractor:function}, its limitations in handling complex \textit{Memory} operations can result in incomplete or inaccurate extraction~\cite{grech2019gigahorse, zhang2023bian}.

\vspace{-0.1in}
\begin{algorithm}
\small
\captionsetup{font={small}}
\caption{Data-flow-based heuristic arguments recovery}
\label{algo:extract}
\SetAlFnt{\ttfamily} 
\KwIn{\textit{CS}, the callsite of a function call}
\KwOut{\textit{Args}, argument positions and values}
$\textit{PC} \leftarrow \textit{FreeMemPointer(CS)} $

$\textit{base} \leftarrow \textit{GetBase(PC)} $

$\textit{idx} \leftarrow 0, \textit{Args} \leftarrow []$

\For{\textit{pc} from \textit{PC} to \textit{CS}}{
    \If{$\textit{isMSTORE(pc)}$}{
        $\textit{(ptr, Var)} \leftarrow \textit{ParseMSTORE(pc)}$
        
        $\textit{offset} \leftarrow \textit{ptr - base}$
        
        $\textit{value, flag} \leftarrow $ \textit{DataFlowRecover(Var)}

        \If{$!(\textit{offset = 0x4}$ \textbf{and} $\textit{idx = 0})$}{

            \If{$flag = True$}{
                $\textit{err} \leftarrow \textit{TypeCheck(idx, value)}$
                
                \If{$err \ != \emptyset$}{
                    $ReportError(err)$
                    
                    \textbf{continue} 
                }
   
            }

             \Else{
               $\textit{value} \leftarrow $ \textit{GetSymbolicValue()}
            }  
        
            $\textit{Args.append((value, idx))}$ 
        }
        $\textit{idx += 1}$
    } 
}
\Return $\textit{Args}$
\end{algorithm}
\vspace{-0.1in}

To address this, we present a \textit{data-flow-based heuristic arguments recovery algorithm}.
The EVM manages \textit{Memory} using a free memory pointer, typically designating the offset at position \texttt{0x40} as the pointer to the next available memory location~\cite{sollang, pan2023automated}. 
Based on this heuristic, Algorithm~\ref{algo:extract} presents the overview of how we extract arguments from \textit{Memory}.
Specifically, given a function callsite, we backtrace to find the nearest \texttt{MLOAD} instruction that loads a value from \texttt{0x40} (L1) and consider the loaded value as the $base$ for arguments (L2). Then, we traverse all \texttt{MSTORE} instructions between the \texttt{MLOAD} and the callsite and extract the target position ($ptr$) and the to-be-stored variable ($Var$) (L4 -- L6). To this end, we can extract the offset of the variable by subtracting $base$ from $ptr$ (L7), and calculate the concrete value of the variable through data flow analysis (L8). If this step fails due to complex control flow, a unique symbolic value is assigned to the argument to maintain its positional information (L16). However, due to stack operations and untyped \textit{Memory} accesses in the EVM~\cite{li2024varlifter}, the algorithm may violate our assumptions, resulting in mismatches between the recovered values and their expected types. Therefore, we handle errors based on type checking (L11 -- L14), where we compare the recovered argument types at the same index with those declared in the function declaration. For example, if the type of the argument calculated based on the offset is identified as \texttt{address} with the length of \texttt{0x20}, but should be \texttt{uint256} with the length of \texttt{0x32} in the function declaration, the types are considered mismatched. Then an error is reported, and the recovery process for the argument is skipped. 
Consequently, the remaining function arguments are identified by their index positions and returned (L17 and L19).

\subsubsection{High-Level Semantics Combining} 
\label{lab:semantics}
Current recovered information does not reveal the specific behavior of function calls. To identify attack intentions, we need to incorporate high-level semantics to extract token-related operations within the contract. We propose two complementary approaches, \textit{i.e.,} 1) leveraging function signature templates and 2) utilizing argument positional analysis. 

As for the first approach, we heuristically consider that attackers frequently interact with Flashloan and DEXes to construct attack chains. To identify such interactions, we utilize function signature templates associated with widely adopted token standards, including ERC20~\cite{erc20}, ERC721~\cite{erc721}, and ERC1155~\cite{erc1155}. Moreover, we extract and validate the semantics of functions from commonly used Flashloan and DEXes services based on total value locked (TVL) and transaction volume~\cite{defillama}, \textit{e.g.,} Aave~\cite{aave}, Pancake~\cite{pancakeswap}, and Uniswap~\cite{uniswap}, by referring to their official documentation.

\begin{figure}[t] 
\centering
\includegraphics[width=\columnwidth]{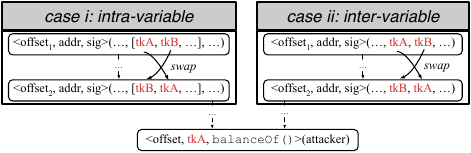} 
\vspace{-0.2in}
\caption{Heuristically infer potential token swap actions.} 
\vspace{-0.1in}
\label{fig:pswap} 
\end{figure}

If no templates are matched, we take another heuristic, \textit{i.e.,} the order of involved tokens would be swapped in price manipulation. Therefore, we adopt an \textit{argument positional analysis} to identify potential token swaps for cases where function signature templates are unavailable, as illustrated in Figure~\ref{fig:pswap}.
Specifically, we consider whether the extracted xFCG has the following features.
There exist two nodes with identical callee addresses and function selectors, while the order of any two address-type variables is reversed. The reverse can happen on the intra-variable (case \textbf{\textit{i}}) or inter-variable (case \textbf{\textit{ii}}) level. We heuristically consider these two addresses to be involved in a token swap.
As a rational attacker, confirming whether obtaining profits is necessary.
We heuristically take the invocation to \texttt{balanceOf()} of a swapped address as necessary in the xFCG.
If both above rules meet in the xFCG, we reckon there exists a token swap of the involved two addresses.

Consequently, within all collected templates as well as the identified token swap behaviors, we have summarized five token action related semantics into consideration, \textit{i.e.,} transfer, flashloan, and liquidity-related operations, whose formal definitions are shown in Figure~\ref{fig:syntax-dsl}:

\begin{itemize}
\item \textbf{Transfer (\textit{Tr}).} Transfer a specified amount ($amt$) of a $token$ from one address ($from$) to another one ($to$).
\item \textbf{Swap token (\textit{ST}.}) In a liquidity pool ($pr$), swap an amount of input token ($tk\_in,\ amt\_in$) to output token ($tk\_out,\ amt\_out$), which is sent to an address ($to$).
\item \textbf{Add liquidity (\textit{AL}).} Against a liquidity pool ($pr$), deposit some token ($amt\_in,\ tk\_in$) to mint some LP token ($amt\_out,\ tk\_out$) to an address ($to$).
\item \textbf{Remove liquidity (\textit{RL}).} Against a liquidity pool ($pr$), burn some LP token ($amt\_in,\ tk\_in$) to transfer some token ($amt\_out,\ tk\_out$) to an address ($to$).
\item \textbf{Flashloan (\textit{FL}).} A flashloan $pr$ lends a specific amount ($amt$) of $token$ to an account ($to$).
\end{itemize}

\subsubsection{Token Flow Graph Construction} 

With the identified semantics, we construct the token flow graph (TFG) based on the xFCG to analyze the contract behavior. It helps guide the identification of critical execution paths in attacks. To formally define TFG, we begin by the following notations:
\begin{itemize}
    \item $\mathcal{N}$, the set of nodes in TFG, representing function calls along with their recovered arguments and semantics.
    \item $\mathcal{E}$, the set of edges in TFG, $\mathcal{E}  \subseteq \mathcal{N} \times \mathcal{N} $, corresponding to the control flow or data dependencies among nodes.
    \item $\mathcal{A}$, the token action related semantics label defined in \S\ref{lab:semantics}.
    \item $\mathcal{T}:\mathcal{N} \rightarrow \mathcal{A}$, a mapping function from nodes to the corresponding semantics labels.
\end{itemize}

\begin{figure}[t] 
\centering
\includegraphics[width=0.90\columnwidth]{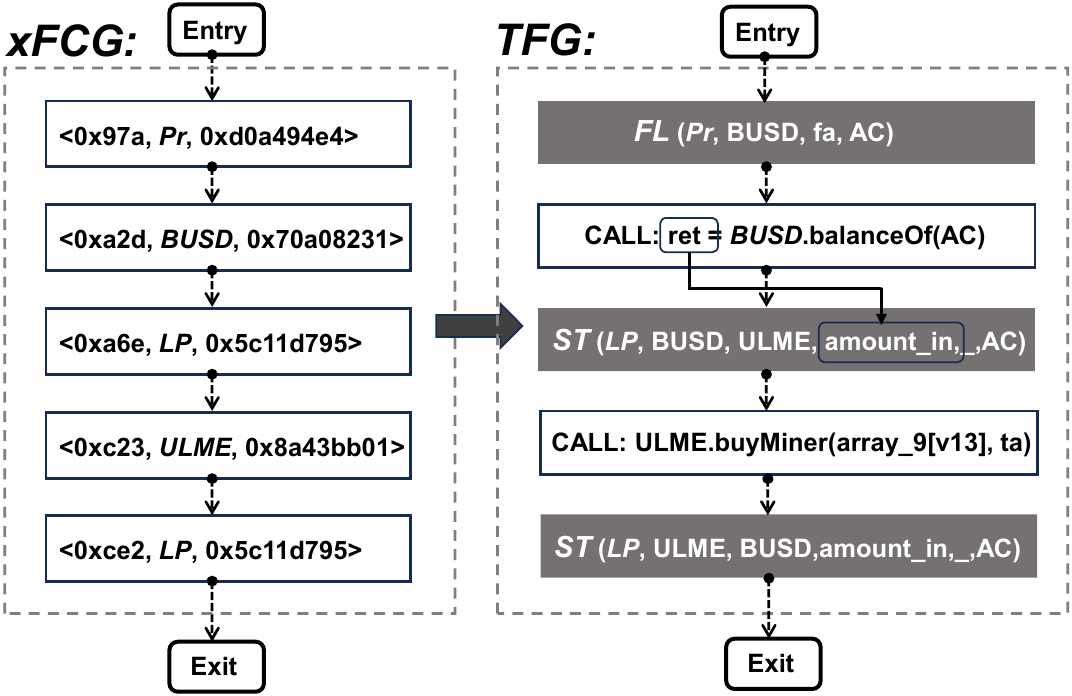} 
\vspace{-0.1in}
\caption{The xFCG and TFG of the attack contract in the \textit{ULME} incident, where AC denotes the attack contract, and the dotted and solid lines refer to control- and data-flow dependencies, respectively.}
\vspace{-0.1in}
\label{fig:tfg4} 
\end{figure}

Token flow graph is defined as $\mathcal{G} = (\mathcal{N}, \mathcal{E}, \mathcal{A}, \mathcal{T})$.
Figure~\ref{fig:tfg4} shows the generated TFG of the \textit{ULME} incident introduced in \S{\ref{sec:moti_eg}}.
As we can see, we retain the control flow dependency relations in the original xFCG and extend nodes with their corresponding token action related semantics. 
Additionally, we parse the data flow dependencies among nodes, as the solid line in Figure~\ref{fig:tfg4}, representing the result of \texttt{balanceOf} is used for swapped amount.
Integrating data flow analysis enables us to precisely track how token-related data propagates, offering insights into the dependencies and effects of each token action. For instance, this allows us to trace whether a user manipulates the liquidity pool with borrowed funds following a Flashloan.

\begin{figure}[t!]
\centering
\resizebox{\columnwidth}{!}{
\begin{minipage}{\columnwidth}
\begin{alignat}{4}
    &  \langle \mathit{addr}   \rangle	\quad 	&&::\!&&= && \quad  \textbf{addresses}  \nonumber\\
    &  \langle \mathit{arg}   \rangle	\quad 	&&::\!&&= && \quad  \textbf{consts} \ | \ \textbf{vars}   \nonumber\\
    &  \mathit{Op}	\quad 	&&::\!&&= && \quad  \mathit{ST} \ | \ \mathit{AL} \ | \ \mathit{RL}  \nonumber\\
	&  \mathit{transfer}  \quad 	&&::\!&&= &&  \quad \mathit{Tr}\ (\text{token:}  \ \langle \mathit{addr}   \rangle, \ \text{from:} \ \langle \mathit{addr}   \rangle, \ \text{to:} \  \langle \mathit{addr}   \rangle,   \nonumber\\ 
    & && && && \quad \quad \text{amt:} \ \langle \mathit{arg} \rangle) \nonumber \\
    &  \mathit{flashloan}  \quad 	&&::\!&&= &&  \quad \mathit{FL}\ (\text{pr:}  \ \langle \mathit{addr}   \rangle, \ \text{token:} \ \langle \mathit{addr}   \rangle, \ \text{amt:} \  \langle \mathit{arg}   \rangle,   \nonumber\\
    & && && && \quad  \quad \text{to:} \ \langle \mathit{addr} \rangle) \nonumber \\
    &  \mathit{liquidity}  \quad 	&&::\!&&= &&  \quad \mathit{Op}\ (\text{pr:}  \ \langle \mathit{addr}   \rangle, \ \text{tk\_in:} \ \langle \mathit{addr}   \rangle, \ \text{tk\_out:} \  \langle \mathit{addr}   \rangle,   \nonumber\\
    & && && && \quad  \quad \text{amt\_in:} \ \langle \mathit{arg} \rangle,\ \text{amt\_out:} \ \langle \mathit{arg} \rangle,\ \text{to:}  \ \langle \mathit{addr} \rangle ) \nonumber
\end{alignat}
\end{minipage}
}
\vspace{-0.1in}
\caption{The definition of token action related semantics.}
\vspace{-0.1in}
\label{fig:syntax-dsl}
\end{figure}

\subsection{Anomalous Token Flow Detector}
To improve the detection efficiency, we perform cross-contract analysis as necessary. Thus, we propose a \textit{sensitive path filtering} method to avoid getting stuck in those meaningless and recursive paths.
Moreover, we design a set of formal rules to robustly identify both direct and indirect price manipulation behaviors introduced in \S\ref{sec:pm}.

\subsubsection{Sensitive Path Filtering}
\label{sec:spf}
As stated in \textbf{Challenge 2}, performing cross-contract analysis for every external call is highly time-consuming and contradicts the timeliness required for identifying attack contracts. Therefore, we only initiate cross-contract analysis on those \textit{suspicious sensitive paths}.
Based on previous empirical studies and analysis against DeFi attacks~\cite{wu2023defiranger, su2021evil, zhou2023sok}, a DeFi attack typically consists of three stages, \textit{i.e.,} fund preparation, token exchange, and fund transfer. Hence, when traversing TFG, cross-contract analysis will only be conducted on the corresponding nodes.
For a clear illustration, we define the following notations:
\begin{itemize}
    \item \textit{Entry} and \textit{Exit} represent the starting and ending nodes of TFG, respectively, to facilitate analysis.
    \item \textit{p} = ($\textit{N}_{1}$, $\textit{N}_{2}$, ..., $\textit{N}_{\text{k}}$), a $k$-tuple to represent a path in TFG, where $N_{i} \in \mathcal{N}$. For convenience, we adopt $N \in p$ to represent the node $N$ is included in the path $p$. $\textit{N}_{1} \prec \textit{N}_{2}$ indicates that $\textit{N}_{1}$ is a predecessor of $\textit{N}_{2}$ in a path.
\end{itemize}

We use $\mathcal{S}_{\textit{p}}$ to represent the set of sensitive paths and $\sigma$ denotes the sender address or contract address.
Let \textit{p} = ($\textit{N}_{\text{start}}$, ..., $\textit{N}_{\text{end}}$), $\textit{p} \subseteq \mathcal{S}_{\textit{p}}$ once if any the following conditions holds:

\begin{itemize}
    \item Fund preparation: $\mathcal{T}$($\textit{N}_{\text{start}}$) = $\textit{FL}$(\_,\_,\_,$\sigma$)  $\wedge$ $\textit{N}_{\text{end}}$ = \textit{Exit}.

    \item Token exchange: $\mathcal{T}\text{(}\textit{N}_{\text{start}}\text{)}$ = $\textit{ST}$(\_,\_,\_,\_,\_,\_) $\wedge \mathcal{T}\text{(}\textit{N}_{\text{end}}\text{)}$ = $\textit{ST}$(\_,\_,\_,\_,\_,\_).

    \item Fund transfer: $\textit{N}_{\text{start}}$ = \textit{Entry}$\wedge$$\mathcal{T}$($\textit{N}_{\text{end}}$) = \textit{Tr}(\_,\_,$\sigma$,\_).
    
\end{itemize}

The complexity of the naive implementation of identifying sensitive paths is $O(n^2)$. Thus, to improve efficiency, we only keep the \textit{longest path}. That is, if $\forall N\in p, N\in p'$ and $p' \subseteq \mathcal{S}_{\textit{p}}$, only $p'$ will be kept. The cross-contract analysis will be conducted on external calls in $p'$.
For example, because \texttt{buyMiner()} in Figure~\ref{fig:tfg4} lies on both the fund-preparation and token-exchange sensitive paths, we will perform the cross-contract analysis to the ULME contract.

\subsubsection{Detecting Rules}
\label{subsubsec:detecting rules}
According to the definition of price manipulation, as we illustrated in \S\ref{sec:pm}, we can identify the attack behavior based on the following rules.

\noindent
\textbf{Rule 1: Pump-and-Dump.}
We first identify if there are two \textit{ST} actions interacting with an identical liquidity pool and drain and deposit tokens to potentially manipulate the token price. 
As shown in Rule~\ref{rule:st}, we use the predicate $\textit{P}_{\textit{PD}}$ to represent the path $p$ along with the two related \textit{ST} actions that hold this property.
Note that, $a \rightarrow a'$ indicates there is a data flow dependency relationship from $a$ to $a'$.
\begin{small}
\begin{equation}
\label{rule:st}
\frac{
    \begin{array}{c}
    \textit{N}_{1} \in \textit{p}, \textit{N}_{2} \in \textit{p}, \textit{N}_{1} \prec \textit{N}_{2}, \textit{p} \subseteq \mathcal{S}_{\textit{p}} \\
    \text{$\mathcal{T}$($\textit{N}_{1}$) = \textit{ST}(\textit{pr}, \_, \textit{t}, \_, $a$, \_)}, \text{$\mathcal{T}$($\textit{N}_{2}$) = \textit{ST}(\textit{pr}, \textit{t}, \_, $a'$, \_, $\sigma$)} \\
    a \rightarrow a'
    \end{array}
}{
    \text{$\textit{P}_{\textit{PD}}$(\textit{p}, $\textit{N}_{1}$, $\textit{N}_{2}$)}
}
\end{equation}
\end{small}

\noindent
\textbf{Rule 2\&3: Direct Price Manipulation.}
Based on Rule~\ref{rule:st}, we can formally define how direct price manipulation is identified. As shown in Rule~\ref{rule:dpm1}, we adopt the predicate \textit{DPM} to capture the price of tokens ($t$ and $t'$) of which the liquidity pool ($pr$) is manipulated.
Such direct price manipulation should be completed by taking advantage of flashloan services, where the borrowed tokens will be directly used in $N_1$.

\begin{small}
\begin{equation}
\label{rule:dpm1}
\frac{
    \begin{array}{c}
    \text{$\textit{P}_{\textit{PD}}$(\textit{p}, $\textit{N}_{1}$, $N_2$), $\textit{N}_{0}$ $\in$ \textit{p}, $\textit{N}_{0}$ $\prec$ $\textit{N}_{1}$} \\
    \text{$\mathcal{T}$($\textit{N}_{0}$) = \textit{FL}(\textit{t}, \_, $\sigma$, $a$)}, \text{$\mathcal{T}$($\textit{N}_{1}$) = \textit{ST}(\textit{pr}, \textit{t}, \textit{t'}, $a'$, \_, $\sigma$)} \\
    a \rightarrow a'
    \end{array}
}{
    \text{\textit{DPM}(\textit{pr}, \textit{t}, \textit{t'})}
}
\end{equation}
\end{small}

Except for Rule~\ref{rule:dpm1}, Rule~\ref{rule:dpm2} also demonstrates a type of direct price manipulation. Their distinction exists in whether a victim is involved (the entity shown in Figure~\ref{fig:elephant} in \S\ref{sec:dpm}).
If a victim exists, there will be another action between the two in $P_{PD}$. As shown in Rule~\ref{rule:dpm2}, the victim should be involved in a token swap between itself and the pool or add / remove liquidity to / from the pool to manipulate the token price.
Finally, the $DPM$ predicate records the victim address ($vc$) and both involved tokens.
\begin{small}
\begin{equation}
\label{rule:dpm2}
\frac{
    \begin{array}{c}
    \text{$\textit{P}_{\textit{PD}}$(\textit{p}, $\textit{N}_{1}$, $\textit{N}_{3}$), $\textit{N}_{2}$ $\in$ \textit{p}, $\textit{N}_{1}$ $\prec$ $\textit{N}_{2}$ $\prec$ $\textit{N}_{3}$ } \\
    \text{$\mathcal{T}$($\textit{N}_{1}$) = \textit{ST}(\textit{pr}, \textit{t}, \textit{t'}, \_, $a$, \_)} \\
    \text{$\mathcal{T}$($\textit{N}_{2}$)=\textit{Tr}(\textit{t}, \textit{vc}, \textit{pr},\_)$\vee$\textit{AL}(\textit{pr},\textit{t},\_,\_,\_,\textit{vc})$\vee$\textit{RL}(\textit{pr},\_,\textit{t'},\_,\_,\textit{vc})}
    \end{array}
}{
    \text{\textit{DPM}(\textit{vc}, \textit{t}, \textit{t'})}
}
\end{equation}
\end{small}

\noindent
\textbf{Rule 4: Indirect Price Manipulation.}
As for indirect price manipulation, we adopt $IPM$ to capture its characteristics, whose formal definition is shown in Rule~\ref{rule:ipm1}.
As we can see, it looks similar to Rule~\ref{rule:dpm2}. The difference is located on whether the price fluctuation is caused by the victim address ($vc$) or the attacker itself ($\sigma$). 
\begin{small}
\begin{equation}
\label{rule:ipm1}
\frac{
    \begin{array}{c}
    \text{$\textit{P}_{\textit{PD}}$(\textit{p}, $\textit{N}_{1}$, $\textit{N}_{3}$), $\textit{N}_{2}$ $\in$ \textit{p}, $\textit{N}_{1}$ $\prec$ $\textit{N}_{2}$ $\prec$ $\textit{N}_{3}$ } \\
    \text{$\mathcal{T}$($\textit{N}_{1}$) = \textit{ST}(\textit{pr}, \textit{t}, \textit{t'}, \_, $a$, \_)} \\
    \text{$\mathcal{T}$($\textit{N}_{2}$)=\textit{Tr}(\_, \textit{vc}, $\sigma$,\_)$\vee$\textit{AL}(\textit{vc},\textit{t},\_,\_,\_,$\sigma$)$\vee$\textit{RL}(\textit{vc},\_,\textit{t'},\_,\_,$\sigma$)}
    \end{array}
}{
    \text{\textit{IPM}(\textit{vc}, \textit{t}, \textit{t'})}
}
\end{equation}
\end{small}

\section{Implementation \& Experimental Setup}
\label{sec:evaluation}

\noindent
\textbf{Dataset.}
To comprehensively evaluate {\tool}, we have collected two datasets, \textit{i.e.,} a ground-truth dataset ($\mathcal{D}_{G}$) and a large-scale contracts dataset ($\mathcal{D}_L$).
Specifically, $\mathcal{D}_G$ consists of two sub-datasets.
$\mathcal{D}_{G1}$ comprises 84 attack events labeled as price manipulation, sourced from various mainstream platforms~\cite{Defihacklab, slowmist, rekt}, the publicly released datasets of FlashSyn~\cite{chen2024flashsyn} and DeFiRanger~\cite{wu2023defiranger}, and manually verified by two of our authors specializing in DeFi security\footnote{Seven attack incidents labeled by FlashSyn and DeFiRanger are excluded. Please refer to Table~\ref{tab:not_selected} in Appendix for the reasons.}.
Moreover, we heuristically take the number of transactions generated by a contract as the criterion to determine its benign nature. Thus, we use APIs from Etherscan~\cite{EthApi} and data from TokenTerminal~\cite{tokenter} to select the top 8,000 active contracts by transaction volume as non-malicious cases ($\mathcal{D}_{G2}$).
$\mathcal{D}_L$ encompasses all contracts deployed in the recent two years, from April 2022 to June 2024, covering over 770K contracts in total.
We deploy an Ethereum archive node using Geth~\cite{geth} to replay transactions and collect them.
Finally, to demonstrate {\tool}'s capability in real-time detection, we define the \textit{attack time window} as the period from the contract's deployment to the initiation of a price manipulation attack that results in a profitable transaction against the victim contract.

\noindent \textbf{Baseline Selection.}
To the best of our knowledge, there is no tool that supports detecting price manipulation attack contracts based solely on the contract bytecode.
To evaluate the effectiveness of {\tool}, we select three the most relevant state-of-the-art tools as baselines, \textit{i.e.,} DeFiRanger (DR)~\cite{wu2023defiranger}, FlashSyn (FS)~\cite{chen2024flashsyn}, and DeFiTainter (DT)~\cite{kong2023defitainter}.
Because DR and FS are close-sourced and detect price manipulation based on transaction data, we directly take the results from their papers.
Moreover, DT is an open-source tool that can detect the potential victims of price manipulation on the bytecode level. Thus, we provide all victims contract bytecode of incidents in $\mathcal{D}_{G1}$ to evaluate its effectiveness.
Note that though DeFiGuard~\cite{wang2024defiguard} claims it can extract behavioral features from transactions and use graph neural networks to identify price manipulation attacks, its model is close-sourced and it does not release the corresponding dataset. Thus, we exclude DeFiGuard from baselines.

\noindent \textbf{Implementation.}
Based on the facts and IR generated from Gigahorse~\cite{grech2019gigahorse}, {\tool} is implemented in Python3, comprising 1.8K lines of code. Additionally, {\tool} utilizes custom declarative rules to obtain more detailed data flow and call stack information, implemented through 500 lines of Datalog. 
All our experiments are conducted on a 96-core server equipped with dual Intel(R) Xeon(R) Gold 6248R CPUs and 256GB RAM running Ubuntu 22.04.1 LTS. The timeout for Gigahorse decompilation is set to 120 seconds.
The recursive cross-contract analysis depth in SMARTCAT is configurable and is set to three in the following evaluation.

\noindent \textbf{Research Questions.} We aim to explore the following research questions (RQs):

\begin{itemize}

\item[\textbf{RQ1}]  Is {\tool} effective and robust in identifying price manipulation attack contracts on the bytecode level?

\begin{itemize}
\item[\textbf{RQ1.1}] What about the performance improvements of {\tool} over baselines?
\item[\textbf{RQ1.2}] Do the introduced methods, \textit{i.e.,} argument recovery algorithm and sensitive path filtering module, contribute positively to the final results?
\item[\textbf{RQ1.3}] How robust is SMARTCAT against obfuscation?
\end{itemize}

\item[\textbf{RQ2}] How many price manipulation attack contracts exist in the wild? What are their characteristics?

\item[\textbf{RQ3}] Can {\tool} be taken as a real-time detector?

\end{itemize}

\section{RQ1: Effectiveness and Robustness}
\label{sec:rq1}

\begin{table}[t]
\caption{Comparison of detecting results on $\mathcal{D}_G$ among {\tool} and baselines. ST represents our tool {\tool}, ST w/o R, ST w/o S, and ST w/o B excludes the argument recovery algorithm module, the sensitive path module, and both of them, respectively.}
\vspace{-0.1in}
\centering
\label{tab:results}
\resizebox{\columnwidth}{!}{%
\begin{tabular}{@{}ccccccccc@{}}
\toprule
\multirow{2}{*}{\textbf{Metrics}} & \multirow{2}{*}{\textbf{\#Detect}} & \multicolumn{5}{c}{\textbf{$\mathcal{D}_{G1}$}}                                                                    & \multicolumn{2}{c}{\textbf{$\mathcal{D}_{G2}$}} \\ \cmidrule(l){3-9} 
                                  &                                    & \textbf{TP} & \textbf{FN} & \textbf{Recall}   & \textbf{Time (s)} & \multicolumn{1}{c|}{\textbf{\#Alert}} & \textbf{FP}   & \textbf{Precision}       \\ \midrule
\textbf{ST}                       & \textbf{79}                        & \textbf{77} & \textbf{7}  & \textbf{91.7\%} & \textbf{32.18}   & \textbf{68}                           & \textbf{2}    & \textbf{99.975\%}   \\
\multicolumn{1}{l}{\textbf{ST w/o R}}                   & 49                                 & 47          & 37          & 56.0\%         & 28.79            & 40                                    & 2             & 99.975\%            \\
\multicolumn{1}{l}{\textbf{ST w/o S}}                   & 83                                 & 77          & 7           & 91.7\%          & 86.36            & 52                                    & 6             & 99.925\%            \\
\multicolumn{1}{l}{\textbf{ST w/o B} }                   & 53                                 & 47          & 37          & 56.0\%         & 79.59            & 30                                    & 6             & 99.925\%            \\ \midrule
\textbf{DR}                       & 23                                 & 23          & 19          & 54.8\%         & N/A              & N/A                                   & N/A           & N/A                \\
\textbf{FS}                       & 9                                  & 9           & 12          & 42.9\%         & N/A              & N/A                                   & N/A           & N/A                \\
\textbf{DT}                       & 60                                 & 14          & 70          & 16.7\%         & N/A              & N/A                                   & 46            & 99.425\%            \\ \bottomrule
\end{tabular}%
}
\vspace{-0.1in}
\end{table}

To answer RQ1, we apply {\tool} and other baselines on $\mathcal{D}_G$ to quantitatively evaluate their effectiveness. We also conduct an ablation study to evaluate the contribution of the argument recovery algorithm and sensitive path filtering to the final results.
Finally, we assess {\tool}'s robustness by applying it to obfuscated smart contracts.

\subsection{RQ1.1: Comparison with Baselines}
\noindent 
\textbf{Overall Results.}
Table~\ref{tab:results} illustrates the overall results of {\tool} and the other three baselines on $\mathcal{D}_G$\footnote{The detailed results are in Table~\ref{tab:detailed_results} in Appendix.}.
Because both DR and FS are close-sourced, we can only evaluate their performance according to the results in their papers.
According to their data, out of 42 and 21 incidents, DR and FS only detect 23 and 9 ones, respectively.
As for DT, it can only detect 14 vulnerable contracts out of 84 cases in $\mathcal{D}_{G1}$. We observe that DT cannot deal with recent attack incidents (see Table~\ref{tab:detailed_results} in Appendix). We speculate the reason is that it relies on manually crafted expert knowledge, while these new cases integrate more advanced and complicated business logic.
In contrast, {\tool} successfully identifies 77 price manipulation attack contracts and demonstrates its effectiveness on recent attack incidents.
In other words, in terms of recall, {\tool} (91.7\%) outperforms the other three baselines (54.8\%, 42.9\%, and 16.7\%) significantly.
As for the efficiency, the average detection time of {\tool} is only 32.18 seconds. According to the results, {\tool} can alert the attack within the attack time window for 68 out of 84 cases (81.0\%). For the remaining 16 cases, 7 are due to detection failures, while the other 9 have an attack window of less than 20 seconds, which poses further challenges to detection efficiency and also points out our future research direction.

As for $\mathcal{D}_{G2}$, among 8,000 benign contracts, {\tool} only generates 2 false positives. DT, however, produces 46 false positives. A manual review confirmed that these contracts do not contain vulnerabilities detected by DT.
This is because DT only considers the token balance of external addresses as a taint source without accounting for constraints such as slippage protection or maximum swap limits, leading to a higher rate of false positives.
Overall, {\tool} demonstrates a higher precision (99.975\% vs. 99.425\%) compared to the state-of-the-art baseline.

\noindent \textbf{False Negative Analysis}.
We manually investigate seven false negatives in $\mathcal{D}_{G1}$ and summarize three root causes.
First, {\tool} relies on the decompilation results of Gigahorse, which have inherent limitations. As its authors said~\cite{grech2019gigahorse}, Gigahorse cannot decompile all valid Ethereum contracts. For example, case \#1\footnote{Indexed in Table~\ref{tab:detailed_results} in Appendix, same notations hereafter.} is written in Vyper~\cite{vyper}, another valid programming language for Ethereum contracts but not widely-adopted, which is not supported by Gigahorse.
Moreover, Gigahorse cannot correctly decompile case \#51 even if we extend the timeout to 60 minutes.
Second, {\tool} depends on accurately recovering function calls and arguments. In cases \#19 and \#48, the attack contracts adopt obfuscation techniques in MEV bots~\cite{mevObfus}, dynamically passing offset values from \textit{Calldata} and calculating function selectors with predefined magic numbers, which invalidate {\tool}. 
In case \#18, the function arguments involve complex dynamic types or custom structures, which render our heuristic argument recovery algorithm ineffective, leading to the failure to correctly identify the semantics.
Third, {\tool} does not consider those attacks that require multiple transactions. In case \#23, the attacker completes the attack through two transactions, \textit{i.e.,} calling \texttt{stake()} to deposit and exchange tokens and then calling \texttt{harvest()} to execute the attack. Case \#28 is similar. {\tool} fails to detect these attacks because it focuses on those attack contracts that embed their logic within a single transaction for rapid exploitation.
We further discuss these limitations in \S\ref{sec:discuss}.

\begin{figure}[h]
\begin{lstlisting}[caption={The \texttt{swapTokenForFund()} function.}, label=lst:fpcode, belowskip=-2em]
function swapTokenForFund(uint256 tokenAmount) private {
    path[0] = address(this);
    path[1] = usdt;
    _swapRouter.swapExactTokensForXXX(amount1, 0, path, tokenDistributor);
    uint256 usdtBalance = USDT.balanceOf(tokenDistributor);
    USDT.transferFrom(tokenDistributor, address(this), usdtBalance);
    ...
    uint256 rewardUsdt = usdtBalance-fundUsdt-lpUsdt;
    if (rewardUsdt > 0 && usdt != _rewardToken) {
        path[0] = usdt;
        path[1] = _rewardToken;
        _swapRouter.swapExactTokensForXXX(rewardUsdt,0,path,address(this));}
}

\end{lstlisting}
\end{figure}

\noindent
\textbf{False Positive Analysis.}
As for the two false positives in $\mathcal{D}_{G2}$, further investigation reveals that both contracts implement the \texttt{swapTokenForFund()} function, as shown in Listing~\ref{lst:fpcode}.
Its token flow aligns with the indirect price manipulation behavior (step \textbf{i} to \textbf{iii} in \S\ref{sec:ipm}).
Specifically, the function first executes a swap operation to exchange for USDT tokens (L2-L4), which might affect its price in the liquidity pool (step \textbf{i}). It then transfers USDT from a third-party address to the contract (L5), potentially staking the tokens at an unfairly calculated price (step \textbf{ii}). Finally, the function calculates the reward tokens to be returned to the caller based on a predefined fee ratio (L7-L10) and returns a portion of the USDT to the liquidity pool (L12) (step \textbf{iii}). Consequently, {\tool} mistakenly identifies it as an attack contract.
The original intent of this function is to distribute the incoming \texttt{tokenAmount} by converting it to USDT, allocating portions to designated addresses, managing liquidity, and finally swapping any remaining USDT back to the contract.
Though we can add extra rules to eliminate such false positives, it may lead to other unexpected false negatives. As a detector specifically designed for identifying attack contracts with timely alerts, we choose to accept false alarms to mitigate possible attacks more proactively.

\subsection{RQ1.2: Ablation Study}
We perform an ablation study by removing the argument recovery algorithm (see \S\ref{lab:argu}) and the sensitive path filtering module (see \S\ref{sec:spf}) to evaluate their contributions.
As shown in Table~\ref{tab:results}, ST w/o R can only detect 47 attack contracts out of 84 cases in $\mathcal{D}_{G1}$, mainly due to its inability to accurately identify token actions for specific function calls, resulting in an incomplete TFG. This means that integrating the argument recovery algorithm introduces 63.8\% more true positive cases, while only introducing the runtime overhead of 3.4 seconds.
As for ST w/o S, the number of detected attacks on $\mathcal{D}_{G1}$ is consistent with {\tool}. However, removing it dramatically increases the average detection time by 1.7$\times$, \textit{i.e.,} 16 attacks cannot be alerted in time. 
Moreover, on $\mathcal{D}_{G2}$, ST w/o S introduces 4 additional false positives. 
We find that these four cases invoke the two false positives identified by {\tool}. Since all execution paths are treated as potentially sensitive paths, the cross-contract operations are also incorrectly considered part of the attack. This demonstrates that the sensitive path filtering module helps our tool focus on suspicious paths, significantly reducing cross-contract analysis time.
Intuitively, ST w/o B, which does not integrate both modules, performs poorly in terms of efficiency, precision, and recall, underlining the significance of integrating these two modules in {\tool}.

\subsection{RQ1.3: Robustness}

We further evaluate the robustness of {\tool} against code obfuscation.
Currently, two mature obfuscators for Ethereum smart contracts are available, \textit{i.e.,} BOSC~\cite{yu2022bytecode} and BiAn~\cite{zhang2023bian}.
As BOSC performs on the deployed bytecode but does not guarantee the deployability after obfuscation, we adopt BiAn as the obfuscator.
To be specific, BiAn performs source-level obfuscation with three modes: Layout Obfuscation (LAO), Data Flow Obfuscation (DFO), and Control Flow Obfuscation (CFO). We note that 1) the \texttt{replaceVarName} option of LAO changes external interface definitions, causing mismatched function selectors of cross-contract calls; and 2) the maintainer has also acknowledged that CFO is not yet functional~\cite{bian}. Thus, to ensure the contract functionality, we only consider LAO, DFO, and LDO (combined with LAO and DFO) as the obfuscation methods, consistent with a previous work~\cite{chen2021sadponzi}.
Since most attack contracts are only available as bytecode, we construct a dataset of 20 contracts by \textit{i.e.,} 1) reverse-engineering the bytecode of detected attack cases to reproduce their source code; and 2) modifying PoCs of attacks reported by security platforms~\cite{Defihacklab}.
We use Foundry~\cite{foundry} to deploy the obfuscated contracts on a forked private chain, simulating on-chain real-time detection scenarios. Note that the environment is reinitialized before testing each case to prevent caching from disturbing the final results.
As shown in Table~\ref{tab:robustness}, {\tool} accurately identifies all contracts obfuscated by all three modes with acceptable runtime overhead. Such robustness can be explained by two factors, \textit{i.e.,} 1) LAO's variable name replacement does not affect bytecode analysis; and 2) Gigahorse provides a robust data flow analysis, while {\tool} focuses on semantic information that remains unchanged during DFO.
We further discuss conducting obfuscation on attack contracts in \S\ref{sec:discuss}.

\begin{table}[t]
\centering
\caption{Performance of {\tool} under obfuscation.}
\label{tab:robustness}
\vspace{-0.1in}
\resizebox{0.7\columnwidth}{!}{%
\begin{tabular}{ccccc}
\toprule
\textbf{Mode}    & \textbf{None} & \textbf{LAO} & \textbf{DFO} & \textbf{LDO} \\ \midrule
\textbf{FN}      & 0 / 20          & 0 / 20         & 0 / 20         & 0 / 20         \\ 
\textbf{Avg. Time (s)} & 45.1      & 45.8         & 47.2         & 47.5         \\ 
\bottomrule
\end{tabular}%
}
\vspace{-0.1in}
\end{table}

\begin{tcolorbox}[title= Answer to RQ1, left=2pt, right=2pt, top=0.5pt,bottom=0.5pt, colback=gray!5,colframe=gray!80!black]
Based on our comprehensive dataset, {\tool} outperforms all state-of-the-art baselines, effectively identifying price manipulation attack contracts with 91.6\% recall and $\sim$100\% precision. Additionally, it can alert 81.0\% of attack contracts within the attack window and demonstrate robustness against obfuscation techniques. The ablation study confirms the significant role of the argument recovery algorithm and sensitive path filtering module.
\end{tcolorbox}

\section{RQ2: Real-world Price Manipulation}

To answer RQ2, we apply {\tool} on over 770K contracts in $\mathcal{D}_L$, and characterize the financial impacts of identified attack contracts. Moreover, we also quantify the efficiency of {\tool} on a large-scale experiment.

\noindent
\textbf{Overall Results.}
In total, we have identified 616 price manipulation attack contracts, none of which are linked with source code on Etherscan.
We utilized auxiliary information, like transaction traces and account labels to confirm the detection results. Specifically, we tracked whether transactions involved in any swap token operations and profited from price fluctuations. We also take advantage of labels on Etherscan~\cite{EthApi} and reports from security service platforms~\cite{Defihacklab, slowmist, rekt}. 

Among them, till May 2024, 214 have already launched the corresponding attack, accounting for 34.7\%, where 19 of them were reported in public. {\tool} can promptly raise alerts within the attack window for 195 cases (91.1\%), with an average detection time of 27.6 seconds.
Interestingly, out of 214 cases, we have investigated 40
failed attack contracts and 8 successful but no-profit contracts. After investigating their decompiled bytecode and transactions, we identified two primary reasons, \textit{i.e.,} 1) insufficient prerequisite conditions, such as inadequate funds to destabilize liquidity pools or lack of access control permissions, and 2) changes in on-chain states that caused attackers to miss profit opportunities.

For the remaining 402 cases without initiating attack transactions, to further analyze the effectiveness of our tool, we randomly sampled 20 contracts. Taking advantage of an online tool~\cite{dedaub}, we obtained the decompiled code of these contracts for examination. Inspired by previous studies~\cite{ren2024lookahead, forta}, we also consider flashloan services, deployer information, bytecode length, and function signatures. Ultimately, among the 20 sampled contracts, we confirmed 16 as price manipulation attack contracts as they demonstrated clear attack intent, such as implementing functions related to flashloan callback, asset transfer, and interaction with liquidity pool. The remaining 4 are inconclusive because either their code is obfuscated or there is no additional information about the deployer. We speculate there are two reasons why so many attack contracts remain in the pre-attack stage. On the one hand, to ensure timeliness and profitability, many attack contracts may deploy attack templates in advance, allowing attackers to initiate exploitation at any time by passing in the address of the target liquidity pool. On the other hand, we found delayed attacks in $\mathcal{D}_{G1}$, where one case occurred 44 days after deployment. Thus, we speculate attackers may have missed the opportunity or are still waiting for the ripe time to maximize their profits.

\begin{figure}[t]
    \centering
    \subfigure[Distribution of profits for reported and unreported attack contracts.]{
        \centering
        \includegraphics[width=0.95\columnwidth]{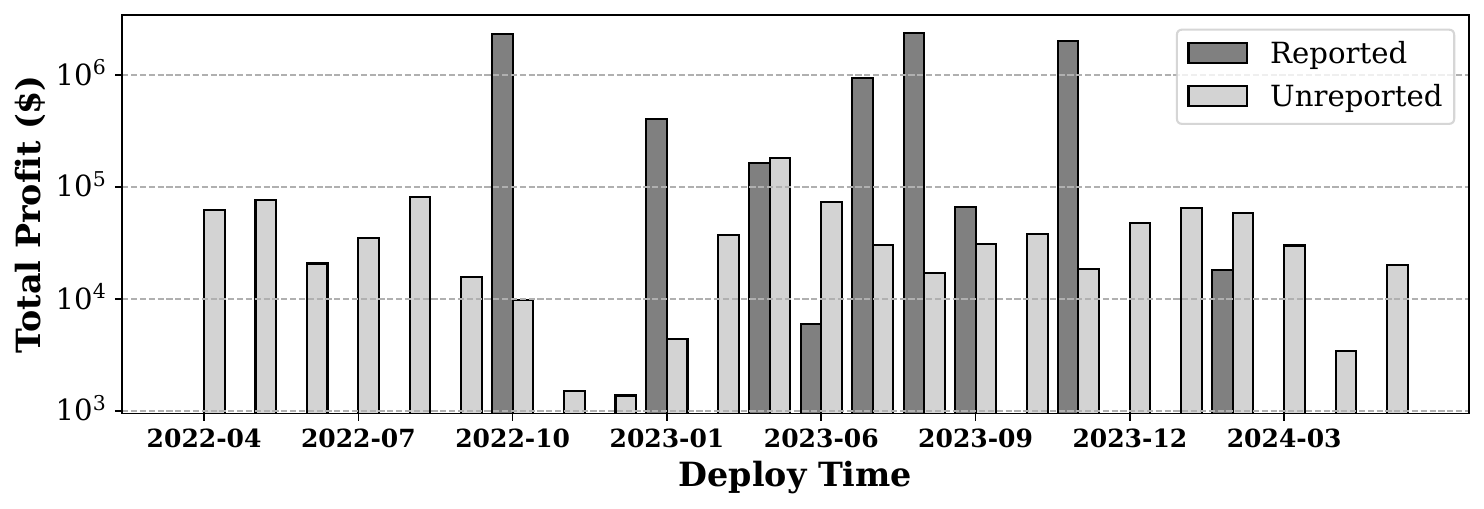}
        \label{fig:reported}
    }
    \subfigure[Distribution of profits for two types of price manipulation attack contracts, with numbers on bars indicating contract count.]{
        \centering
        \includegraphics[width=0.95\columnwidth]{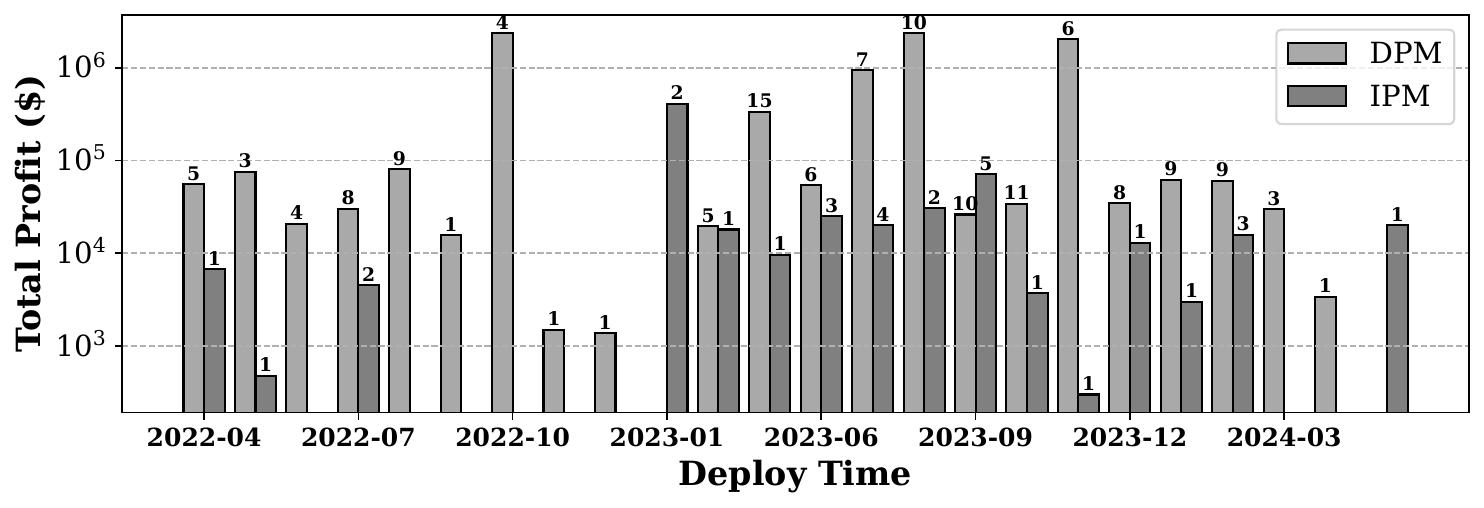}
        \label{fig:pmtype}
    }
    \vspace{-0.1in}
    \caption{Distribution of attack profits by deployment time.}
    \vspace{-0.1in}
    \label{fig:profit}
\end{figure}

\noindent
\textbf{Financial Impacts.} 
In total, attackers have obtained \$9.25M in profits through 166 price manipulation attack contracts\footnote{214 cases without 40 failed and 8 no-profit cases.}, out of which \$0.96M are not reported at all (related to 147 successful contracts).
Though the reported attacks have accounted for 89.6\% of financial losses, we cannot neglect the impact of unreported ones.
Figure~\ref{fig:reported} illustrates the distribution of attack profits.
As we can see, price manipulation attacks have persisted alongside the growth of Ethereum, indicating that attackers have consistently targeted potential vulnerabilities in the DeFi ecosystem with the intent to steal funds. There is a noticeable trough in late-2022, which may be linked to the collapse of the FTX project and the increased scrutiny by the U.S. Securities and Exchange Commission on crypto institutions~\cite{ftx}. This has led to a more conservative approach to the development of the DeFi ecosystem. It reflects that the activity level of price manipulation attacks is closely related to the growth and evolution of DeFi.
Additionally, Figure~\ref{fig:pmtype} illustrates the number and profitability of the two types of price manipulation attacks. In total, 136 direct price manipulation (DPM) attack contracts generated a profit of \$8.60M, far exceeding the \$0.65M from 30 indirect price manipulation (IPM) attack contracts. This disparity arises because attackers often use flashloans to directly manipulate the number of tokens in liquidity pools, causing price fluctuations. In contrast, IPM attacks interact with third-party addresses, requiring more effort and costs for attackers.

\begin{figure}[t] 
\centering
\includegraphics[width=0.95\columnwidth]{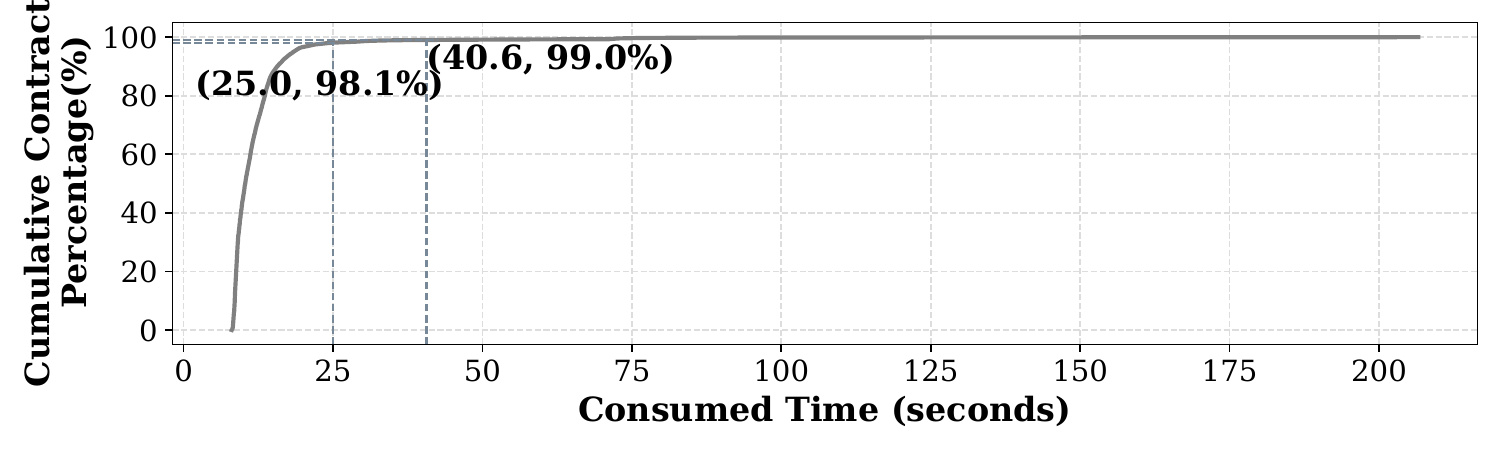} 
\vspace{-0.1in}
\caption{Distribution of contract runtimes.} 
\vspace{-0.1in}
\label{fig:cdf} 
\end{figure}

\begin{figure}[t] 
\centering
\includegraphics[width=0.95\columnwidth]{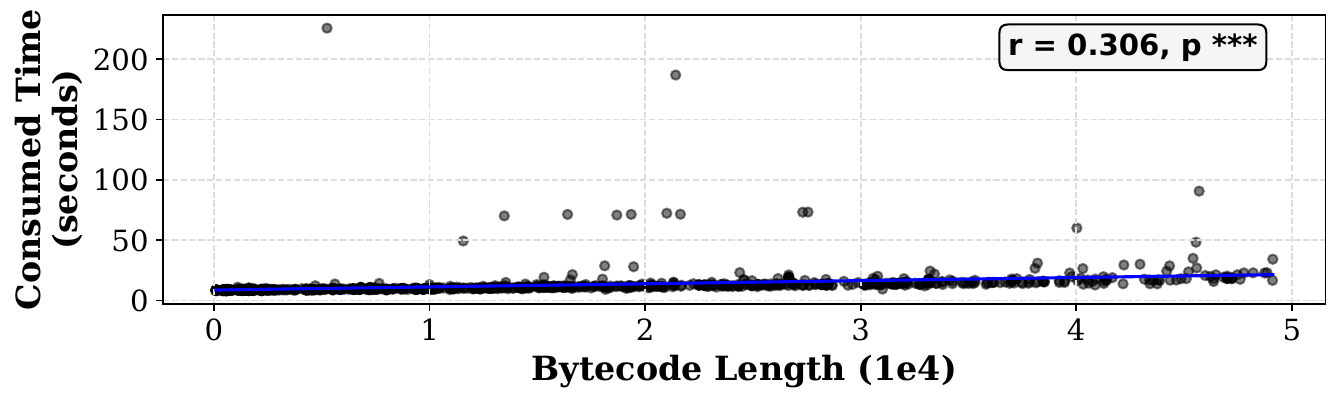} 
\vspace{-0.1in}
\caption{The relationship between the consumed time and the length of the bytecode on each case. } 
\vspace{-0.1in}
\label{fig:byteCode_Time} 
\end{figure}

\noindent
\textbf{Efficiency.}
We further analyze the efficiency of {\tool} on $\mathcal{D}_L$. We sampled 15K contracts from $\mathcal{D}_L$, assigning a separate process to each contract for analysis. The distribution of the used time is shown in Figure~\ref{fig:cdf}. 
As we can see, 98.1\% contracts can be finished within 25 seconds. If we extend the runtime to 40.6 seconds, 99\% of the contracts can be covered.
Our statistics further show that the average time for a contract is only 12.1 seconds.
Figure~\ref{fig:byteCode_Time} illustrates the relationship between the consumed time and the length of the bytecode based on randomly sampled 1,000 contracts.
We can observe that there is no exponential relationship between these two metrics. This is because {\tool} only performs cross-contract analysis on nodes within sensitive paths during execution (see \S{\ref{sec:spf}}), thereby reducing the time overhead and improving its scalability.
For the top-left outliner that takes 226 seconds\footnote{Contract address: 0xA5C0D0CAf243697143ed9f06b259050A77cE5887}, we find that {\tool} does not perform additional cross-contract analysis. Instead, 158 seconds are spent on Gigahorse inlining small functions to produce a higher-level IR, and 65 seconds are used for generating facts for client execution.
The correlation analysis only illustrates the weak linear relationship between them, where $r$ is 0.306 with $p < 0.001$, further proving the efficiency of {\tool}.

\begin{tcolorbox}[title= Answer to RQ2, left=2pt, right=2pt, top=0.5pt,bottom=0.5pt, colback=gray!5,colframe=gray!80!black]
{\tool} has identified 616 price manipulation attack contracts in total, accounting for \$9.25M in financial losses, where only 19 cases were reported publicly.
Moreover, {\tool} can analyze 98.1\% of cases within 25 seconds, and there is only a weak linear relationship between the consumed time and bytecode length ($r = 0.306$ with $p < 0.001$), demonstrating the efficiency and scalability of {\tool} on real-world tasks.
\end{tcolorbox}

\section{RQ3: Real-time Detection}
To answer RQ3, we have deployed {\tool} on Ethereum and Binance Smart Chain (BSC) as a real-time detector since July 11th, 2024.
We monitor the latest blocks using Geth RPC nodes~\cite{geth} and extract contract bytecode from contract creation transactions. To accelerate the analysis, we have deployed 15 instances in parallel. Additionally, the number of deployed instances can be dynamically scaled based on the volume of newly created contracts.

\begin{table*}[t]
\caption{Successfully conducted price manipulation attacks alerted by {\tool}. All times are in 2024 and are presented in UTC. Numbers in parenthesis are time windows since deployment in seconds.}
\label{tab:rq3}
\vspace{-0.1in}
\resizebox{\textwidth}{!}{%
\begin{tabular}{@{}clcccccc@{}}
\toprule
\textbf{Victim DApp} & \multicolumn{1}{c}{\textbf{Type}} & \multicolumn{1}{c}{\textbf{Deploy Time}} & \multicolumn{1}{c}{\textbf{Alert Time}} & \multicolumn{1}{c}{\textbf{Attack Time}} & \multicolumn{1}{c}{\textbf{Loss (\$)}} & \multicolumn{1}{c}{\textbf{Attack Transaction Hash}}               \\ \midrule
UPS (BSC)   &   IPM      & 07/12 11:09:12                           & 07/12 11:10:32 (80)                          & 07/12 11:12:24 (192)                                                          & 521K                                  & 0x1ddf415a4b18d25e87459ad1416077fe7398d5504171d4ca36e757b1a889f604 \\
TokenStake (BSC)  & DPM & 08/05 18:52:25                           & 08/05 18:55:02 (157)                          & 08/05 19:11:49 (1,164)                                                        & 110K                                  & 0x94ff0c3f3177a6ffd3365652ae2dc1f0a4ecf5f5758df1fdc3339303992a2ae4 \\
FXS (BSC)    &   DPM   & 08/21 11:56:43                           & 08/21 11:57:21 (38)                          & 08/21 11:58:04 (81)                                                          & 10K                                   & 0xa479ae7d0b53ec8049de7f4556aa9b1d406f51dacd027ebe60f9f45788b7deb2 \\ \bottomrule
\end{tabular}%
}
\vspace{-0.2in}
\end{table*}

\begin{table}[]
\caption{Failed and unfinished price manipulation attempts detected by {\tool}. Same notations with Table~\ref{tab:rq3}.}
\label{tab:rq3-2}
\vspace{-0.1in}
\resizebox{\columnwidth}{!}{%
\begin{tabular}{@{}cclccc@{}}
\toprule
\textbf{Status}                                                                    & \textbf{Address}  & \textbf{Type} & \textbf{Deploy Time} & \textbf{Alert Time}  & \textbf{Chain} \\ \midrule
\multirow{4}{*}{\textbf{\begin{tabular}[c]{@{}c@{}}Failed \\ (4)\end{tabular}}}    & 0xDd02...        & DPM & 07/26 06:50:35       & 07/26 06:52:05 (90)  & BSC            \\
                                                                                   & 0xE39b...  & IPM      & 08/02 18:37:06       & 08/02 18:37:32 (26)  & BSC            \\
                                                                                   & 0x5dB0...  & IPM      & 08/16 16:26:11       & 08/16 16:28:02 (111) & ETH            \\
                                                                                   & 0xbc3E... & DPM      & 08/25 08:14:38       & 08/25 08:16:43 (125) & BSC            \\ \midrule
\multirow{7}{*}{\textbf{\begin{tabular}[c]{@{}c@{}}Unfinished\\ (7)\end{tabular}}} & 0x0EBD...        & DPM & 07/17 05:41:52       & 07/17 05:42:39 (47)  & BSC            \\
                                                                                   & 0x7707... & DPM        & 08/05 19:17:58       & 08/05 19:19:56 (118) & BSC            \\
                                                                                   & 0xd270... & DPM       & 08/13 11:15:59       & 08/13 11:16:50 (51)  & ETH            \\
                                                                                   & 0x2F6C...  & DPM      & 08/13 15:26:45       & 08/13 15:31:14 (269) & BSC            \\
                                                                                   & 0x85Ea...  & DPM      & 08/16 08:16:47       & 08/16 08:17:49 (62)  & BSC            \\
                                                                                   & 0x02F8...  & DPM      & 08/17 09:01:02       & 08/17 09:03:53 (171) & BSC            \\
                                                                                   & 0x35Be...  & DPM      & 08/21 11:59:55       & 08/21 12:00:36 (41)  & BSC            \\ \bottomrule
\end{tabular}%
}
\vspace{-0.1in}
\end{table}

\noindent
\textbf{Overall Results.}
In total, {\tool} has reported 14 cases, as shown in Table~\ref{tab:rq3} and Table~\ref{tab:rq3-2}, illustrating the completed attacks and failed/unfinished ones, respectively.

As we can see from Table~\ref{tab:rq3}, three successful cases were all performed on BSC.
Compared to their corresponding attack window, we can conclude that {\tool} is efficient enough to identify contracts' semantics and raise alarms for the upcoming price manipulation attack.
According to our statistics, these three cases have resulted in financial losses worth more than \$641K.

Moreover, Table~\ref{tab:rq3-2} illustrates four failed and seven unfinished price manipulation attempts.
We manually investigated the bytecode and transaction of four failed attempts. We found that three of them reverted due to an inability to repay the flashloan, and one was due to running out of gas as a result of multiple transfer loops within the transaction.
As for all seven unfinished ones, we have observed explicit attack intent, as the ones we stated in the \textbf{Overall Results} part in RQ2.
Interestingly, we have noticed that the unfinished case located at \texttt{0xd270} has another contract (\texttt{0xb7f2}) with identical bytecode that is reported by Etherscan. We have noticed that \texttt{0xb7f2} is involved in a real-world price manipulation attack\footnote{Attack transaction hash: \href{https://etherscan.io/tx/0x758efef41e60c0f218682e2fa027c54d8b67029d193dd7277d6a881a24b9a561}{link}} which is identified in RQ2. The attack was launched 110 days after the contract deployment and led to around \$1.1M in financial losses. \texttt{0xd270} is deployed six minutes after the attack. 
Thus, we infer that \texttt{0xd270} is either another try from the same team and waiting for the ripe time, or a test contract used by the victim to analyze the cause of the attack.

\begin{lstlisting}[caption={Attack contract against UPS.}, label=lst:atkcode]
// Step_0: use flashloan to borrow BUSD token
function pancakeV3FlashCallback(uint256 varg0, uint256 varg1, bytes _) public nonPayable {
    v1 = BUSD.balanceOf(address(this));
    // Step_1: swap BUSD to UPS
    v2, v3 = sto_2.swapExactTokensForTokens(v1, 1, [BUSD, UPS], address(this), _);
    v4 = UPS.balanceOf(address(sto_0));
    // Step_2: Get UPS token from another contract
    UPS.transferFrom(sto_0, address(this), v4);
    v5 = UPS.balanceOf(address(sto_5)); 
    // Calculate swap amount for price manipulation
    v6 = (v5 - 1) * 20 / 19;
    // step_3: swap UPS to BUSD at a very low price
    v7, v8 = sto_2.swapExactTokensForTokens(v6, 1, [UPS, BUSD], address(this), _);
    BUSD.transfer(sto_9, varg1 + 1000);
    v9 = BUSD.balanceOf(address(this));
    BUSD.transfer(msg.sender, v9);
}
\end{lstlisting}

\noindent
\textbf{Case Study.}
To better illustrate the effectiveness of {\tool}, we conducted a case study of the attack against UPS on BSC (first data row in Table~\ref{tab:rq3}), which has led to \$521K financial losses.
Since the contract does not release its source code, we use an online tool~\cite{dedaub} to obtain its decompiled representation, as shown in Listing~\ref{lst:atkcode}. We perform some necessary simplifications to clearly demonstrate the attack.

The attacker first borrows a large amount of BUSD tokens through a flashloan and then executes the attack logic in \texttt{pancakeV3FlashCallback()} (L2). The attacker swaps the borrowed BUSD tokens to UPS tokens and transfers them to the attack contract itself (L6). Another bunch of UPS tokens are transferred from a third-party address to the attack contract (L9). Finally, the attacker swaps all UPS tokens back to BUSD and transfers them to the call sender (L15 -- L19). This matches the pattern of indirect price manipulation (see \S\ref{sec:ipm}), thus {\tool} raises the alarm and marks the UPS token as the manipulated token.

\begin{lstlisting}[caption={Vulnerable \texttt{\_update()} in UPS.}, label=lst:vulcode]
function _update(address from, address to, uint256 amount) internal virtual override {
    if (inSwapAndLiquify || whiteMap[from] || whiteMap[to] || !(from == pairAddress || to == pairAddress)) {
        super._update(from, to, amount);
        ...
    } else if (to == pairAddress) {
        uint256 fee = amount * 5 / 100;
        if (!inSwapAndLiquify) {
        // Vulnerable point.
            _swapBurn(amount - fee);}}
    ...
}
\end{lstlisting}

To analyze the root cause of this attack, we have tracked to a customized \texttt{\_update()} function in the UPS's \texttt{transferFrom()} function, whose source code is shown in Listing~\ref{lst:vulcode}.
Specifically, it calculates the number of UPS token needed to burn based on an externally provided \texttt{amount} (L9). This operation affects the contract's reserve calculations, thereby manipulating the token exchange price. 
In this case, the attacker passed a predetermined number of UPS tokens, as shown at L13 in Listing~\ref{lst:atkcode}. This operation has led to the reservation of UPS token dropping to 1 during the second token exchange.
As a result, the attacker exchanges a large amount of BUSD by taking advantage of indirect price manipulation.

\begin{tcolorbox}[title= Answer to RQ3, left=2pt, right=2pt, top=0.5pt,bottom=0.5pt, colback=gray!5,colframe=gray!80!black]
Within 50 days, {\tool} has raised 14 alarms about potential price manipulation attacks on Ethereum and BSC 99 seconds after the corresponding contract deployment on average. Notably, these attacks have already led to \$641K financial losses, and seven of them are still waiting for their ripe time.
\end{tcolorbox}
\section{Discussion}
\label{sec:discuss}

In this section, we will discuss some limitations of our work.

Firstly, {\tool} does not consider path feasibility, which may lead to false positives when raising alarms.
However, deploying contracts requires gas fees, and attackers generally aim for profit. It is unlikely that attackers would introduce dead loops or unreachable code in their attack contracts. Therefore, we assume that attackers maintain full control over their contracts and ensure all branches are reachable to execute the attack effectively.

Secondly, {\tool} utilizes function signature templates combined with heuristic rules to extract token flow related semantics in \S\ref{lab:semantics}, which is unable to infer those uncommon function signatures. However, attack contracts typically rely on well-established and widely-used DEXes, which lowers the chances of such oversights. Additionally, integrating large language models in future work could help in identifying the semantics and parameter information of these external calls more accurately as illustrated in~\cite{wang2024smartinv}.

Thirdly, {\tool} relies on accurate decompilation and argument recovery, where complex obfuscation techniques may cause Gigahorse to fail. Attackers may also employ unknown evasion techniques to bypass detection. Since attackers rarely disclose their obfuscation methods, existing research in this area is limited. However, attack contracts are typically one-time-use and lack reusability, making obfuscated attack contracts rare. Moreover, adopting obfuscation incurs extra gas costs, further reducing attackers' willingness to use them. Additionally, {\tool} may struggle to recover arguments correctly when dealing with complex dynamic types or custom structures. Due to the EVM stack and low-level operations, recovering function arguments and types remains an open challenge, as highlighted by VarLifter~\cite{li2024varlifter}.

Another limitation is that {\tool} does not account for attacks conducted across multiple transactions. However, it is intuitive that attackers prefer quick attacks executed within a single transaction to avoid state changes in the victim contracts.
We admit that some attackers would perform deployment and attack within the same block, resulting in an extremely short attack window. Due to the current time constraints of {\tool} in detecting attacks, it may fail to issue timely alerts for such cases, as it relies on Gigahorse for decompiling bytecode, which is time-consuming. 
In the future, we plan to accelerate analysis by simulating the EVM stack directly and filtering out long or irrelevant paths that do not match attack patterns for faster detection.

\section{Related work}
\label{sec:related}

\noindent \textbf{Smart Contract Vulnerability Detection.} Various studies and tools have been proposed to detect hidden vulnerabilities in smart contracts to prevent asset loss for users. Static analysis tools like Slither~\cite{feist2019slither}, SAILFISH~\cite{bose2022sailfish},  Securify~\cite{tsankov2018securify}, VETEOS~\cite{liveteos}, and Ethainter~\cite{brent2020ethainter} analyze source code or bytecode. Gigahorse~\cite{grech2019gigahorse, grech2022elipmoc}  and MadMax~\cite{grech2018madmax} offer a decompilation framework that translates bytecode to a custom IR. AVVERIFIER~\cite{sun2024all} performs taint analysis by simulating the process of EVM stack execution. Symbolic execution tools like Mythril~\cite{mythril}, Oyente~\cite{luu2016making}, ETHBMC~\cite{frank2020ethbmc}, EOSAFE~\cite{he2021eosafe} and Manticore~\cite{mossberg2019manticore} explore potentially vulnerable paths using constraint solvers. Dynamic analysis approaches, such as fuzzing tools~\cite{jiang2018contractfuzzer, grieco2020echidna, manes2018fuzzing, choi2021smartian, shou2023ityfuzz, wustholz2020harvey, he2019learning, ye2023detecting}, generate random inputs or reorder historical transaction sequences to test contracts. Furthermore, GPTScan~\cite{sun2024gptscan} leverages large language models~\cite{he2024large} to localize vulnerabilities by defining vulnerability scenarios.

\noindent \textbf{Price Manipulation Detection.} Price manipulation attacks have long posed a serious threat to the DeFi ecosystem. Existing tools like DeFiRange\cite{wu2023defiranger}, DeFiGuard~\cite{wang2024defiguard}, and FlashSyn~\cite{chen2024flashsyn} detect such attacks based on transaction information. DeFiTainter~\cite{kong2023defitainter} starts from contract bytecode, using cross-contract taint analysis to explore all execution paths and locate entry points of vulnerable functions.

\noindent \textbf{Attack Contract Identification.} Attackers increasingly prefer to launch attacks through contracts. Su \textit{et al.}~\cite{su2021evil} collect key threat intelligence related to DApp attack incidents through measurements and implement an automated tool for large-scale discovery of attack incidents. Forta~\cite{forta} and Lookahead~\cite{ren2024lookahead} extract statistical features of attack contracts and train machine learning models to develop classifiers. Yang \textit{et al.}~\cite{yang2024uncover} construct call chains from attack contracts to victim contracts and locate vulnerable functions based on the characteristics of reentrancy attacks.

\section{Conclusion}
\label{sec:conclu}
In this work, we propose {\tool}, an effective and efficient static analyzer to identify price manipulation attack contracts solely on the bytecode level.
Based on the decompiled intermediate representation, with the help of the data-flow-based heuristic arguments recovery algorithm and sensitive path filtering method, {\tool} successively builds the cross-function callsite graph and token flow graph to characterize the control- and data-flow dependency relationships among function calls.
According to the formally defined rules, {\tool} can achieve 91.6\% recall and $\sim$100\% precision on a ground truth dataset, while also demonstrating robustness against obfuscation techniques. Furthermore, {\tool} has revealed 616 potential price manipulation attack contracts, accounting for \$9.25M financial losses, where only 19 cases were reported publicly. By adopting {\tool} on Ethereum and BSC, {\tool} has raised 14 alarms 99 seconds after the corresponding deployment on average. These alarmed ones have already led to \$641K financial losses, while seven of them are still waiting for their ripe time.

\section*{Ethical Consideration}

In RQ1, both $\mathcal{D}_{G}$ and $\mathcal{D}_{L}$ used in our study are sourced from publicly available blockchain service platforms or social media. The attack contracts in  $\mathcal{D}_{G}$ have already been thoroughly verified by security professionals and are no longer capable of causing further economic harm to the DeFi ecosystem.

During our large-scale analysis of deployed contracts in RQ2, {\tool} successfully identified 616 attack contracts, where only 19 cases were reported previously. We tested these contracts on a private blockchain and confirmed that some of them still have the potential to launch profitable attacks.
We are in the process of contacting the relevant project teams through various channels, including project websites and community platforms. 
Considering that publicly disclosing the addresses of these attack contracts could attract malicious attempts, this part of the data is excluded from our open-sourced dataset. 

In RQ3, during the real-time monitoring of the Ethereum and Binance Smart Chain, {\tool} has successfully raised 14 alerts. Unfortunately, due to the short window between contract deployment and attack execution for the three cases in Table~\ref{tab:rq3}, attackers were able to launch and obtain profits before we could establish contact with the project teams.
As for the seven unfinished ones, we have contacted the relevant project teams once after the alarm is raised.
We strongly encourage stakeholders in the community to integrate tools like {\tool} to prevent or mitigate such kinds of threats. 

\section*{Data Availability}

We have released {\tool} and the ground-truth dataset ($\mathcal{D}_G$) at \url{https://figshare.com/articles/online_resource/SMARTCAT_Artifact/28192028}.

\bibliographystyle{plain}
\bibliography{ref}

\appendix
\section*{APPENDIX}
\label{sec:appendix}

\section{Detailed Detection Results}

\begin{table}[htbp]
\centering
\caption{The attack incidents that are in the released datasets of FlashSyn and DeFiRanger but are not selected.}
\label{tab:not_selected}
\resizebox{\columnwidth}{!}{%
\begin{tabular}{cccl}
\toprule
\textbf{Dataset}      & \textbf{Chain} & \textbf{App}        & \textbf{Reason} \\ 
\midrule
\multirow{5}{*}{FlashSyn}  
                     & ETH           & Eminence            & Design Flaw             \\ 
                     & ETH           & Yearn               & Design Flaw             \\ 
                     & ETH           & bearFi              & Design Flaw             \\ 
                     & BSC           & AutoShark           & Non-Contract             \\ 
                     & BSC           & ElevenFi            & Design Flaw             \\ 
\midrule
\multirow{2}{*}{DeFiRanger} 
                     & ETH           & Dracula             & Undisclosed             \\ 
                     & BSC           & Belt Finance        & Non-Contract              \\ 
\bottomrule
\end{tabular}
}
\end{table}

Table~\ref{tab:not_selected} lists seven attack events that are included in the released datasets of FlashSyn and DeFiRanger but are not selected by this study. We summarize the following three reasons:
\begin{itemize}
 \item\textbf{Design Flaw}:
the incident, which is exclusively the focus of FlashSyn, differs from price manipulation. It arises from flawed logic in one or more functions of the victim contracts, leading to highly specific vulnerabilities.
 \item\textbf{Non-Contract}: the incident is initiated by an EOA instead of a smart contract, which is out of our scope.
 \item\textbf{Undisclosed}: the incident is not reported on any blockchain security platforms/forums or other social media, which cannot 100\% guarantee its existence.
\end{itemize}

\begin{table*}[t]
\caption{Detecting results for different tools on $\mathcal{D}_{G1}$, where \checkmark indicates the corresponding attack/vulnerable contract can be detected. DR, FS, DT, and SC refer to DeFiRanger, FlashSyn, DeFiTainter, and {\tool}, respectively. }
\scriptsize
\centering
\rowcolors{1}{gray!20}{white} 
\label{tab:detailed_results}
\begin{tabularx}{\textwidth}{p{0.1cm}p{1.22cm}p{1.29cm}rp{0.45cm}p{0.3cm}p{0.3cm}p{0.3cm}p{0.3cm}p{0.1cm}p{1.22cm}p{1.29cm}rp{0.45cm}p{0.3cm}p{0.3cm}p{0.3cm}p{0.3cm}}
\toprule
\rowcolor{white}
\textbf{\#} & \textbf{Victim App} & \textbf{Attack Time} & \textbf{Loss} & \textbf{Chain} & \textbf{DR}$^{*} $ & \textbf{FS}$^{*}$ & \textbf{DT}  & \textbf{SC} & \textbf{\#} & \textbf{Victim App} & \textbf{Attack Time} & \textbf{Loss} & \textbf{Chain} & \textbf{DR}$^{*}$ & \textbf{FS}$^{*}$ & \textbf{DT}  & \textbf{SC} \\ 
\midrule
1 & bZx & 2020/02/18 & \multicolumn{1}{r}{350.0K} & ETH & \checkmark  & \checkmark &   &  & 43 & MCC & 2023/05/09 & 19.0K & BSC &  &  &  & \checkmark \\
2 & Balancer & 2020/06/28 & 447.0K & ETH & \checkmark  &  &   &\checkmark & 44 & HODLCapital & 2023/05/09 & 4.3K & ETH &  &  &  & \checkmark \\
3 & Loopring & 2020/09/30 & 29.0K & ETH & \checkmark  &  &    &\checkmark & 45 & SellToken & 2023/05/11 & 95.0K & BSC &  &  &  & \checkmark \\
4 & Harvest & 2020/10/26 & 21.5M & ETH & \checkmark & \checkmark & \checkmark &\checkmark & 46 & LW & 2023/05/12 & 50.0K & BSC &  &  &    & \checkmark \\
5 & PloutoFinance & 2020/10/29 & 650.0K & ETH & \checkmark &  &  &  \checkmark & 47 & CS & 2023/05/23 & 714.0K & BSC &  &  &  & \checkmark \\
6 & CheeseBank & 2020/11/06 & 3.3M & ETH & \checkmark & \checkmark & \checkmark &\checkmark & 48 & EBPools & 2023/05/31 & 111.0K & ETH &  &  &   &   \\
7 & ValueDefi & 2020/11/24 & 6.0M & ETH & \checkmark &  & \checkmark &  \checkmark & 49 & Cellframe & 2023/06/01 &  \multicolumn{1}{r}{76.0K} & BSC &  &  & \checkmark  & \checkmark \\
8 & SealFinance & 2020/11/15 & 4.3K & ETH & \checkmark &  &  &  \checkmark & 50 & VINU & 2023/06/06 & 6.0K & ETH &  &  & & \checkmark \\
9 & WarpFinance & 2020/12/17 & 7.8M & ETH & \checkmark & \checkmark & \checkmark & \checkmark & 51 & UN & 2023/06/06 & 26.0K & BSC &  &  &   &\\
10 & ApeRocket & 2021/07/14 & 1.3M & BSC &  & \checkmark & \checkmark  & \checkmark  & 52 & SellToken & 2023/06/11 & 106.0K & BSC &  &  &  & \checkmark \\
11 & ArrayFinance & 2021/07/18 & 516.0K & ETH & \checkmark &  &  &  \checkmark & 53 & CFC & 2023/06/15 & 16.0K & BSC &  &  &  & \checkmark \\
12 & Zenon & 2021/11/20 & 1.0M & BSC & \checkmark &  &  &   \checkmark & 54 & IPO & 2023/06/20 & 483.0K & BSC &  &  &  & \checkmark \\
13 & CollectCoin & 2021/12/01 & 1.0M & BSC & \checkmark &  &  &  \checkmark & 55 & SHIDO & 2023/06/23 & 230.0K & BSC &  &  & & \checkmark \\
14 & IVM & 2021/12/17 & 1.0M & BSC & \checkmark &  &  & \checkmark & 56 & LUSD & 2023/07/07 & 16.0K & BSC &  &  &  & \checkmark  \\
15 & MIGE & 2022/02/09 & 42 & BSC & \checkmark &  &  & \checkmark & 57 & WGPT & 2023/07/12 &  80.0K & BSC &  &  &  &  \checkmark  \\
16 & OneRing & 2022/03/21 & 1.5M & FTM &  & \checkmark &  & 
 \checkmark & 58 & ApeDAO & 2023/07/18 & 7.0K & BSC&  &  &   & \checkmark  \\
17 & WienerDOGE & 2022/04/25 & 30.0K & BSC &  & \checkmark & \checkmark & \checkmark & 59 & ConicFinance & 2023/07/23 & 934.0K & ETH &  &  &  & \checkmark  \\
18 & bDollar & 2022/05/21 & 2.3K & BSC & \checkmark &  &  &    & 60 & Uwerx & 2023/08/02 & 324.0K & ETH&  &  &  & \checkmark  \\
19 & Novo & 2022/05/29 & 89.6K & BSC & \checkmark & \checkmark &  &   & 61 & Zunami & 2023/08/14 & 2.0M & ETH&  &  &  & \checkmark  \\
20 & Fswap & 2022/06/13 & 432K & BSC & \checkmark &  &  &   \checkmark & 62 & BTC20 & 2023/08/19 & 30.0K & ETH &  &  &  & \checkmark  \\
21 & InverseFi & 2022/06/16 & 1.3M & ETH & \checkmark & \checkmark & \checkmark  &   \checkmark & 63 & EHIVE & 2023/08/21 & 15.0K & BSC &  &  &  &  \checkmark \\
22 & SpaceGodzilla & 2022/07/13 & 25.0K & BSC & \checkmark  &  &  & \checkmark & 64 & GSS & 2023/08/24 & 24.8K & BSC&  &  &  & \checkmark \\
23 & EGDFinance & 2022/08/07 & 36.0K & BSC &  &  & \checkmark &   & 65 & EAC & 2023/08/29 & 6.3K & BSC &  &  &  & \checkmark  \\
24 & ZFinance & 2022/09/05 & 61.0K & BSC &  &  &  &  \checkmark  & 66 & JumpFarm & 2023/09/05 & 4.0K & ETH &  &  &  &  \checkmark  \\
25 & NewFreeDAO & 2022/09/08 & 1.9M & BSC &  &  & \checkmark & \checkmark & 67 & HCT & 2023/09/07 & 6.5K & BSC&  &  &  & \checkmark  \\
26 & BXH & 2022/09/28 & 40.0K & BSC &  &  & \checkmark &  \checkmark  & 68 & UniclyNFT & 2023/09/16 & 6.0K & ETH &  &  &  &\checkmark  \\
27 & TempleDao & 2022/10/11 & 2.3M & ETH &  &  & &  \checkmark & 69  & KubSplit & 2023/09/24 & 78.0K & BSC &  &  &  &   \checkmark \\
28 & ATK & 2022/10/12 & 127.0K & BSC &  &  &  &    & 70 & BH & 2023/10/11 & 1.3M & BSC &  &  &  &  \checkmark  \\
29 & MToken & 2022/10/16 & 1.0M & BSC &  &  & \checkmark  &  \checkmark & 71 & MicDao & 2023/10/18 & 13.0K & BSC &  &  &  &  \checkmark  \\
30 & PlantWorld & 2022/10/17 & 24.5K & BSC &  &  &  & \checkmark & 72 & UniverseToken & 2023/10/27 & 1.5M & BSC &  &  &  &  \checkmark  \\
31 & HEALTH & 2022/10/20 & 8.8K & BSC & \checkmark &  &  & \checkmark & 73 & OnyxProtocol & 2023/11/01 & 2.0M & ETH &  &  &  &  \checkmark   \\
32 & Market & 2022/10/24 & 65.0K & POL &  &  &  &  \checkmark & 74 & 3913Token & 2023/11/02 & 31.3K & BSC &  &  &  &  \checkmark  \\
33 & n00dleswap & 2022/10/25 & 31.0K & ETH &  & & \checkmark  &  \checkmark  & 75 & Grok & 2023/11/10 & 56.0K & ETH &  &  &  & \checkmark  \\
34 & ULME & 2022/10/25 & 50.0K & BSC & \checkmark &  &  & \checkmark & 76 & Token8633 & 2023/11/17 & 52.0K & BSC &  &  &  & \checkmark  \\
35 & VTFToken & 2022/10/27 & 111.0K & BSC &  &  &  & \checkmark  & 77 & ElephantStatus & 2023/12/06 & 165.0K & BSC&  &  &  & \checkmark  \\
36 & UPSToken & 2023/01/18 & 405.0K & ETH &  &  & \checkmark & \checkmark & 78 & BurnsDefi & 2024/02/05 & 67.0K & BSC &  &  &  & \checkmark \\
37 & BscAnt3 & 2023/01/19 & 426K & BSC & \checkmark  &  &  &  \checkmark & 79 & GAIN & 2024/02/21 & 18.0K & ETH &  &  &  &  \checkmark  \\
38 & SHEEP & 2023/02/10 & 3.0K & BSC & &  & & \checkmark  & 80 & WSM & 2024/04/04 & 18.0K & BSC&  &  &  & \checkmark  \\
39 & DYNA & 2023/02/22 & 23.0K & BSC &  &  &  & \checkmark  & 81 & UPS & 2024/04/09 & 28.0K & BSC&  &  &  & \checkmark  \\
40 & AToken & 2023/03/21 & 28.4K & BSC &  &  &  &\checkmark & 82 & MARS & 2024/04/16 & 100.0K & BSC &  &  &  & \checkmark  \\
41 & BIGFI & 2023/03/22 & 30.0K & BSC &  &  &  & \checkmark & 83 & SATX & 2024/04/16 & 27.6K & BSC &  &  &  & \checkmark  \\
42 & SFM & 2023/03/28 & 8.0M & BSC & \checkmark  & &  &  \checkmark  & 84 & Z123 & 2024/04/22 & 135.0K & BSC&  &  &  & \checkmark  \\
\bottomrule
\end{tabularx}
\makebox[\linewidth][l]{\hspace{2em}\scriptsize{$^{*}$ Due to their closed-source nature, DeFiRanger and FlashSyn cannot be tested on incidents occurring after 2023/3/28 and 2022/6/16, respectively.}}
\end{table*}

\end{document}